\documentclass[11pt, a4paper]{article}

\pdfoutput=1

\usepackage[english]{babel}
\usepackage[utf8]{inputenc}
\usepackage{jcappub_ver}
\usepackage{enumerate}
\usepackage{amsbsy}
\usepackage{amsmath} 
\usepackage{graphics}
\usepackage{mathrsfs}
\usepackage{wrapfig}
\usepackage{morefloats}
\usepackage{graphicx}
\usepackage{subcaption}
\usepackage[normalem]{ulem}
\usepackage{mciteplus}          

\mciteSetSublistMode{s}
\mciteSetSublistLabelBeginEnd{\par}{\relax}{\relax}
\mciteSetMidEndSepPunct
  {\ifmciteBstWouldAddEndPunct.\else\fi\space}
  {\ifmciteBstWouldAddEndPunct.\else\fi}
  {\relax}

\newcommand{\bea}{\begin{eqnarray}} \newcommand{\eea}{\end{eqnarray}}
\newcommand{\el}{\nonumber \\}
\newcommand{\re}[1]{(\ref{#1})}

\newcommand{\pat}{\partial}

\renewcommand{\sec}[1]{section \ref{#1}}
\newcommand{\fig}[1]{figure \ref{#1}}

\newcommand{\para}{\paragraph}

\renewcommand{\d}{\delta}

\renewcommand{\l}{\lambda}

\newcommand{\LCDM}{$\Lambda$CDM\ }
\newcommand{\GN}{G_{\mathrm{N}}}
\newcommand{\ha}{\frac{1}{2}}

\newcommand{\keq}{k_{\mathrm{eq}}}

\newcommand{\teq}{t_{\mathrm{eq}}}

\newcommand{\rmd}{\mathrm{d}}

\newcommand{\ie}{i.e.\ }
\newcommand{\eg}{e.g.\ }

\newcommand{\adot}{\dot{a}}
\newcommand{\addot}{\ddot{a}}

\newcommand{\Hdot}{\dot{H}}

\newcommand{\half}{\frac{1}{2}}
\newcommand{\rhob}{\bar{\rho}}
\renewcommand{\in}{\textrm{in}}

\newcommand{\av}[1]{\langle{#1}\rangle}
\newcommand{\sQ}{\mathcal{Q}}
\newcommand{\sR}{{^{(3)}R}}

\newcommand{\Om}{\Omega_{\mathrm{m}}}

\newcommand{\om}{\omega_{\mathrm{m}}}

\newcommand{\ob}{\omega_{\mathrm{b}}}
\newcommand{\OQ}{\Omega_{\sQ}}

\newcommand{\OR}{\Omega_{R}}

\newcommand{\Ov}{\Omega_{\Delta\theta}}

\newcommand{\lei}{\l_{\mathrm{ext},i}}

\title{Evaluating backreaction with the ellipsoidal collapse model}

\author{Francesco Montanari}

\author{and Syksy R\"{a}s\"{a}nen}

\affiliation{University of Helsinki, Department of Physics \\
and Helsinki Institute of Physics \\
P.O. Box 64, FIN-00014 University of Helsinki, Finland}

\emailAdd{syksy.rasanen@iki.fi}

\emailAdd{francesco.montanari@helsinki.fi}

\abstract{
We evaluate the effect of structure formation on the average expansion rate with a statistical treatment where density peaks and troughs are modelled as homogeneous ellipsoids. This extends earlier work that used spherical regions. We find that the shear and the presence of filamentary and planar structures have only a small impact on the results. The expansion rate times the age of the universe $Ht$ increases from 2/3 to $0.83$ at late times, in order of magnitude agreement with observations, although the change is slower and takes longer than in the real universe. We discuss shortcomings that have to be addressed for this and similar statistical models in the literature to develop into realistic quantitative treatment of backreaction.
}

\begin{document}

\begin{flushleft}
	\hfill		 HIP-2017-25/TH \\
\end{flushleft}

\maketitle

\setcounter{tocdepth}{2}

\setcounter{secnumdepth}{3}

\section{Introduction} \label{sec:intro}

\para{Evaluating backreaction.}

There is much observational evidence that the universe is statistically homogeneous and isotropic, with a homogeneity scale of the order 100 Mpc
\cite{Bonnor1986, *Stoeger1987, Clarkson:2010uz, *Heavens:2011mr, *Clifton:2011sn, *Hoyle:2012pb, Hogg:2004vw, *Labini:2009zi, *Labini:2010aj, *Labini:2011tj, *Labini:2011dv, *Scrimgeour:2012wt, *Nadathur:2013mva, *Labini:2014zoa, *Laurent:2016eqo, *Ntelis:2017nrj, Rasanen:2009mg, Maartens:2011yx}. However, this does not imply that the average expansion rate necessarily obeys the Friedmann--Robertson--Walker (FRW) equations, which describe a universe that is exactly homogeneous and isotropic. The effect of clumpiness on average evolution is called backreaction \cite{Ellis:1984bqf, *Ellis:1987zz, Ellis:2005uz, Rasanen:2011ki, Buchert:2011sx}. It has been suggested that backreaction could explain the late-time cosmic acceleration \cite{Buchert:1999mc, Wetterich:2001kr, Schwarz:2002ba, Rasanen:2003fy, *Rasanen:2004sa}. In Newtonian gravity backreaction reduces to a boundary term \cite{Buchert:1995fz}. In general relativity this is not the case \cite{Buchert:1999mc}, but the average expansion rate (though not necessarily the luminosity distance \cite{Enqvist:2009hn}) is nevertheless close to its FRW value if the metric is close to the FRW metric \cite{Rasanen:2011ki} (see \cite{Green:2010qy, *Green:2013yua, *Green:2014aga, *Buchert:2015iva, *Green:2015bma, *Ostrowski:2015pzb, *Green:2016cwo} for a related debate). However, it is difficult to estimate the effect in general relativity without assuming that the metric is everywhere close to the same global FRW metric. It has been shown with an exact toy model that structures can have a large effect on the expansion rate and light propagation, even when they are small and the universe is statistically homogeneous and isotropic \cite{Lavinto:2013exa}, but the magnitude of the effect in the real universe remains unclear.

The problem can be approached in two ways. First, without calculating the effect of structures on the expansion rate, it is possible to formulate consistency tests of the FRW metric \cite{Clarkson:2007pz, Rasanen:2013swa, Rasanen:2014mca} and compare them to observations \cite{Shafieloo:2009hi, *Mortsell:2011yk, *Sapone:2014nna, *Cai:2015pia, *LHuillier:2016mtc, *Yu:2016gmd, *Li:2016wjm, *Wei:2016xti, *Montanari:2017yma, Rasanen:2014mca}. Observations can also be used to test backreaction \cite{Boehm:2013qqa, Lavinto:2013exa}, given a relation between the average expansion rate and light propagation \cite{Rasanen:2008be, Rasanen:2009uw, Clarkson:2011br, Bull:2012zx, Lavinto:2013exa}. Second, we can try to evaluate the effect of the non-linear structures on the expansion rate, either by constructing exact or statistical analytical models (such as models with discrete matter \cite{Clifton:2009jw, *Clifton:2009bp, *Clifton:2012qh, *Clifton:2010fr, *Clifton:2011mt, *Clifton:2013jpa, *Clifton:2014lha, *Clifton:2014mza, *Sanghai:2015wia, *Clifton:2015tra, *Sanghai:2016ucv, *Durk:2016yja, *Bibi:2017urt, *Durk:2017rky, Clifton:2016mxx} and models using data from Newtonian $N$-body simulations
\cite{Wiegand:2010uh, *Wiegand:2011gs, *Roukema:2013cya, *Racz:2016rss, *Roukema:2017doi}), or via relativistic cosmological simulations, some of which are in the weak field regime \cite{Adamek:2013wja, *Adamek:2014gva, *Adamek:2014xba, *Adamek:2015eda, *Adamek:2016zes} (where the effect is known to be small) but others are fully non-linear \cite{Yoo:2012jz, *Bentivegna:2012ei, *Bentivegna:2013ata, *Adamek:2015hqa, *Bentivegna:2016fls} and cosmological \cite{bentivegna:2015flc, *Giblin:2015vwq, *Mertens:2015ttp, *Macpherson:2016ict, *Giblin:2017juu}.

In \cite{Rasanen:2008it} a statistical model was considered, with structures taken to form on peaks and troughs of the initial density field, leading to a distribution of collapsing regions and underdense voids. (This is a simplified picture, not all non-linear structures in the real universe are associated with extrema of the density field \cite{Shandarin:1984, Katz:1993, vandeWeygaert:1994ww, Kofman:1994pz, Porciani:2001er, Ludlow:2010xd}.)
The environment of the extrema was modelled with the Newtonian spherical collapse model and its underdense equivalent. This treatment misses anisotropic structures such as filaments and sheets, which are prevalent features of the real universe \cite{Bond:1995yt, Cautun:2014fwa}.
A related shortcoming, noted in \cite{Rasanen:2008it}, is the absence of internal and external tidal effects, \ie angular shear \cite{Zeldovich:1970, Icke:1973, White:1979, Peebles:1980, Shandarin:1984, Hoffman:1986, Eisenstein:1994ni, vandeWeygaert:1995pz, Audit:1996zj, Bond:1996, DelPopolo:2001fq, Popolo:2002}, which can potentially have a large effect on the average expansion rate. Deviations from spherical symmetry have been found to be important for the calculation of the mass function and expansion rate of collapsing objects \cite{Sheth:1999su, Sheth:2001dp, Ohta:2003zc, Ohta:2004mx, Lam:2008ik, Robertson:2008jr, Angrick:2010qg, Ludlow:2011jx, Paranjape:2012ks, Paranjape:2012jt, Reischke:2016dop}. Mass function calculations are concerned with the distribution of stabilised structures, not the details of the formation process. For backreaction, the opposite is true: the endpoint of structure formation is (at least for collapsing objects) less interesting than the evolution between the linear regime and the final stage.

We extend the calculation of \cite{Rasanen:2008it}, using a Newtonian ellipsoidal model for the structures. For spherically symmetric dust, the average expansion rate is independent of the radial density profile in Newtonian gravity \cite{Buchert:1999pq}, though not in general relativity \cite{Chuang:2005yi, *Paranjape:2006cd, Kai:2006ws, Lavinto:2013exa}. So the Newtonian spherical model covers arbitrary inhomogeneity without anisotropy. In contrast, in order to be tractable, the ellipsoidal model has to be homogeneous, but allows for anisotropy. (Indeed, ``homogeneous anisotropic model'' might be a more appropriate name.)
In this sense, the ellipsoidal model can be seen as a different approximation than the spherical model, not a generalisation of it, although the expansion rate of the spherical case is a subcase of ellipsoidal expansion.
In the context of peaks and troughs, the ellipsoidal model can be thought of as the leading approximation in a series expansion of the gravitational potential around an extremum \cite{Bond:1996}.

In \sec{sec:av} we give the details of the ellipsoidal collapse model and the ensemble averaging that we use. In \sec{sec:res} we give the results for the expansion rate and compare to the spherical case, and consider shear and filamentary and planar structures. In \sec{sec:disc} we discuss the results and how the calculation and similar approaches in the literature would have to be improved in order to go from toy models to realistic quantitative descriptions. In \sec{sec:conc} we summarise our conclusions.

\section{Average quantities and homogeneous ellipsoids} \label{sec:av}

\subsection{Backreaction} \label{sec:Buchert}

\para{The Buchert equations.}

Assuming that matter consists of irrotational dust, but making no other symmetry assumptions, the average volume expansion rate on the hypersurface of constant proper time of observers comoving with the dust is given by the Buchert equations \cite{Buchert:1999er} (see \cite{Buchert:2001sa, Larena:2009md, Rasanen:2009uw} for the case with generalised matter content, including rotation),
\bea
  \label{Ray} 3 \frac{\addot}{a} &=& - 4 \pi\GN \av{\rho} + \sQ \\
  \label{Ham} 3 \frac{\adot^2}{a^2} &=& 8 \pi\GN \av{\rho} - \frac{1}{2} \av{\sR} - \frac{1}{2}\sQ \\
  \label{cons} 0 &=& \av{\rho}\dot{} + 3 \frac{}{} \av{\rho} \ ,
\eea
where $a^3$ is proportional to the proper volume, and $H\equiv\frac{1}{3}\av{\theta}=3\frac{\adot}{a}$ is the average of the local volume expansion rate $\theta$, dot denotes derivative with respect to the proper time of observers comoving with the dust fluid, $\av{}$ is the proper volume average on the hypersurface of constant proper time and $\sR$ is the spatial curvature. The effect of inhomogeneity and anisotropy is quantified by the backreaction variable $\sQ$,
\bea \label{Q}
  \sQ \equiv \frac{2}{3}\left( \av{\theta^2} - \av{\theta}^2 \right) - 2 \av{\sigma^2} \ ,
\eea
where $\sigma^2$ is the shear scalar. For a spherical region treated with Newtonian gravity we have $\sQ=0$, as the positive contribution of the variance and the negative contribution of the shear cancel exactly \cite{Buchert:1999pq}. However, in general, $\sQ\neq0$ for a union of spherical regions, as $\sQ$ is not additive. For a homogeneous ellipsoidal region, the variance is zero and the shear is non-zero, so $\sQ<0$ and a homogeneous ellipsoid expands slower than a corresponding sphere with the same density contrast. As in the spherical case, we can have $\sQ>0$ for a union of such regions if the variance of their average expansion rates is larger than the shear contribution.

\paragraph{The density parameters.}

As in FRW models, the contributions to the expansion rate can be parametrised with relative densities. Dividing \re{Ray} and \re{Ham} by $3 H^2$, we have \cite{Buchert:1999er}
\bea
  \label{q} q &\equiv& - \frac{\Hdot}{H^2} - 1  = \frac{1}{2} \Om + 2 \OQ \\
  \label{omegas} 1 &=& \Om + \OR + \OQ \ ,
\eea
where $q$ is the deceleration parameter, $\Om\equiv 8\pi G_N \av{\rho}/(3 H^2)$, $\OR\equiv-\av{\sR}/(6 H^2)$ and $\OQ\equiv-\sQ/(6 H^2)$ are the density parameters of matter, spatial curvature and the backreaction variable, respectively. It is also useful to define the variance and shear density parameters $\Ov\equiv-(\av{\theta^2}-\av{\theta}^2)/(9H^2)=-\Delta\theta/\av{\theta}^2$ and $\Omega_\sigma\equiv\av{\sigma^2}/(3 H^2)$, so that $\OQ=\Ov+\Omega_\sigma$.
For the sum rules \re{q} and \re{omegas}, it is important that the matter is irrotational dust. This assumption is necessarily violated when collapsing structures stabilise, because turning a collapse around is impossible for irrotational dust. The fact that we consider an ensemble average, not a spatial average, leads to other issues with the density parameters, discussed in \sec{sec:disc}.

\subsection{Evolution of homogeneous ellipsoids} \label{sec:ell}

\para{Dynamical equations.}

We model structures as homogeneous, non-rotating, anisotropic regions described with Newtonian gravity and embedded in a homogeneous and isotropic background, following \cite{Peebles:1980, Bond:1996, Angrick:2010qg}. (For a different treatment of ellipsoidal collapse, see \cite{Audit:1996zj}.) We consider dust matter and a spatially flat background, the evolution of which is described by the Friedmann equations,
\bea
  \label{Fried1} 3 \frac{\dot{\bar{a}}^2}{\bar{a}^2} &=& 8\pi\GN\rhob + \Lambda \el
  \label{Fried2} 3 \frac{\ddot{\bar{a}}}{\bar a} &=& - 4\pi\GN\rhob + \Lambda \ ,
\eea
where $\bar a(t)$ is the background scale factor, $\rhob\propto\bar a^{-3}$ is the background density and $\Lambda$ is the cosmological constant. We also denote $\bar H\equiv\dot{\bar a}/\bar a$.

The gravitational potential $\Phi$ in a single region is related to the density by the Poisson equation,
\bea \label{Poisson}
  \nabla^2 \Phi = 4\pi\GN \rho - \Lambda \ ,
\eea
where the derivatives $\nabla^2$ are with respect to the proper position $r^i$, and $\rho(t)=\rhob+\d\rho$ is the density of the region. The solution of \re{Poisson} that reduces to the background outside the region is\footnote{See \cite{Malament:1995, *Norton:1995} for discussion of the meaning of such a potential in Newtonian theory.}
\bea
  \Phi &=& A(t) + B_i(t) r^i + \ha C_{ij}(t) r^i r^j \ ,
\eea
with $\sum_i C_{ii}=4\pi\GN \rho - \Lambda$. As discussed in \cite{Bond:1996} and mentioned in \sec{sec:intro}, this quadratic expression can be understood as a series expansion around an extremum. (Though we will consider extrema of the density, not the gravitational potential.) The spatially constant term is pure gauge, and the linear term corresponds to uniform translation in space, so we set them to zero. The Poisson equation is elliptic, and boundary conditions are needed to fix the non-trace part of the tensor $C_{ij}$, which describes the deformation. As $C_{ij}$ is symmetric, we can choose coordinates $r^i$ such that $C_{ij}$ is diagonal and write, following the notation of \cite{Angrick:2010qg},
\bea \label{phi}
  \Phi &=& \frac{1}{6} ( 4\pi\GN \rhob - \Lambda ) \d_{ij} r^i r^j + 2\pi\GN \rhob \left( \frac{1}{3} \d + \half b_i \d + \lei \right) \d_{ij} r^i r^j \el
  &\equiv& 2\pi\GN \rhob C_i \d_{ij} r^i r^j - \frac{1}{6} \Lambda \d_{ij} r^i r^j \ ,
\eea
where the first term on the first line corresponds to the background contribution and on the second line we have defined $C_i(a)\equiv\frac{1}{3} (1+\d) + \half b_i \d + \lei$. Here $\d\equiv\d\rho/\rhob$ is the density contrast with respect to the background, $\ha b_i\d$ corresponds to tidal effects due to matter inside the region and $\lei$ to tidal effects due to matter external to the region. The functions $b_i(t)$ are the eigenvalues of the traceless part (so $\sum_i b_i=0$) of the internal tidal contribution to $C_{ij}$,
\bea
  b_i(t) = a_1 a_2 a_3 \int_0^\infty \frac{\rmd y}{ (a_i^2+y) \prod_{j=1}^3 (a_j^2+y)^{1/2} } - \frac{2}{3} \ ,
\eea
where $a_i(t)$ are the dimensionless principal axes of the ellipsoid, normalised so that the physical distance is $r^i(t)=a_i(t) x^i$ (no sum), where $x^i$ is the constant coordinate position of a fluid element. The ratio of the proper volume of the region to the background is $a_1 a_2 a_3/\bar{a}^3$, so $\delta=\bar{a}^3/(a_1 a_2 a_3)-1$.  Correspondingly, $\lei(t)$ are proportional to the eigenvalues of the external part of the tidal tensor, and $\sum_i \lei=0$.

The evolution of $a_i$ is determined by Newton's second law applied to the fluid elements (again, no sum over $i$),
\bea
  \ddot r^i &=& - \frac{\pat \Phi}{\pat r^i} = - 4\pi\GN \rhob C_i r^i + \frac{1}{3} \Lambda r^i\ ,
\eea
from which we get
\bea \label{ait}
  \addot_i &=& - 4\pi\GN \rhob C_i a_i - \frac{1}{3} \Lambda a_i \ .
\eea
From now on, we put $\Lambda=0$. It is convenient to rewrite the time derivatives in \re{ait} in terms of derivatives with respect to $\bar a$. Using $(\dot{\bar{a}}/\bar a)^2\propto\bar{a}^{-3}$ from \re{Fried1}, we obtain
\bea \label{aia}
  a_i'' - \frac{1}{2 a} a_i' + \frac{3}{2 a^2} C_i(a) a_i = 0 \ ,
\eea
where prime denotes derivative with respect to $\bar a$.

In order to solve \re{aia}, we have to specify the external tides parametrised by $\lei(t)$, prescribe how to deal with axes collapsing to zero size, and give the initial conditions. We assume that at early times the region is close to the background, consider only growing modes, and fix the initial conditions at scale factor $a_\in$ with the Zel'dovich approximation \cite{Zeldovich:1970},
\bea \label{aiin}
  a_i(a_\in) = a_\in [ 1 - \l_i(a_\in) ] \ , \quad \ a_i'(a_\in) = 1 - 2 \l_i(a_\in) \ ,
\eea

\noindent where $\l_i$ are the eigenvalues of the tidal tensor (\ie the non-background part of $C_{ij}=\frac{\pat\Phi}{\pat r^i \pat r^j}$) in the linear regime; the normalisation is such that $\sum_i\l_i(a_\in)=\delta_\in$ and the ordering is $\l_1\geq\l_2\geq\l_3$. From \re{phi} we see that as both $b_i$ and $\delta$ are first order small in linear theory, the external tide contribution is required to obtain the correct linear theory result, with $\lei(a)=\frac{a}{a_\in} [\l_i(a_\in)-\frac{1}{3}\delta_\in]$ in the linear regime. Unlike in the spherical case, the environment cannot be neglected if the ellipsoid is embedded in an FRW universe; dropping external tides would correspond to the case of an isolated ellipsoid considered in \cite{Peebles:1980}.
We will consider non-linear evolution, so we need to make further assumptions to define $\lei$ outside of the linear regime.

\para{Non-linear external tides.}

Different prescriptions have been proposed for non-linear external tides. In \cite{Eisenstein:1994ni} they were calculated approximately from a series of spherical shells, neglecting angular distribution of matter within each shell. In \cite{Bond:1996} the following linear and non-linear models were proposed:
\bea
  \label{l} \lei(a) &=& \frac{a}{a_\in} \left[ \l_i(a_\in) - \frac{1}{3} \d(a_\in) \right] \qquad \textrm{linear} \\
  \label{nl} \lei(a) &=& \frac{5}{4} b_i(a) \qquad\qquad\qquad\qquad\quad\ \textrm{non-linear} \ .
\eea
The linear prescription simply extrapolates the functional form of the linear case into the non-linear regime, whereas the non-linear prescription is a simple ansatz that has the same linear behaviour, but is dominated by internal dynamics in the non-linear regime. In \cite{Angrick:2010qg}, the two forms were mixed into the hybrid model, where external tide $\lei$ follows the non-linear model \re{nl} in the direction $i$, unless the axis $i$ turns around, i.e. $\adot_i=0$, in which case the evolution follows the linear model \re{l} after turnaround.

The linear model is unsuitable for our purposes, because we consider underdense as well as overdense regions. If an underdense region does not collapse, its density evolution at late stages is much slower than the extrapolation of the linear regime behaviour (because the density contrast cannot become more negative than $-1$), and the linear tide would dominate and give unphysical results.
We have considered both the non-linear and the hybrid model. For our case, the difference is negligible, and we show results only for the non-linear model.

\para{Virialisation.}

If an ellipsoid starts to collapse along an axis, the axis will shrink to zero in a finite time, leading to infinite density. As the matter is irrotational dust, there is nothing to stop the collapse, so a prescription for virialisation has to be added. In the spherical case considered in \cite{Rasanen:2008it} this did not make much difference, because when all directions collapse at the same time, the proper volume of the structure is small at collapse, and it has negligible effect on the average, as the volume goes to zero faster than the expansion rate diverges. However, when only one direction collapses, the volume of the structure is not necessarily negligible, and indeed filaments and walls occupy a sizeable fraction of the volume of the real universe \cite{Bond:1995yt, Cautun:2014fwa}.

We adopt the prescription based on the tensor virial theorem used in \cite{Angrick:2010qg}. At the instant when the virialisation condition is satisfied for axis $i$, the collapse is stopped and $a_i$ is frozen at the value it has at that point. This means that the acceleration in direction $i$ is momentarily divergent, and energy is not conserved. As we average the expansion rate, not the acceleration, this does not make a difference.

\para{Initial conditions.}

We have now fully defined the dynamical equations for the axis lengths $a_i$. Next we should specify the initial conditions, which in the Zel'dovich approximation reduce to the distribution of $\l_i(a_\in)$; in what follows, we drop the subscript ``$\in$'' when referring to the initial values of $\l_i$ and $\d$. Our regions are a model for nonlinear structure, which we take to form on extrema (peaks and troughs) of the density field, taken to be Gaussian, as expected from inflation and confirmed by observations \cite{Ade:2015xua}.
We parametrise the initial conditions in terms of the peak height (or trough depth) given by the density contrast $\delta=\sum_i\l_i$, and the shape given in terms of the ellipticity $e$ and prolateness $p$, defined as
\bea
  e = \frac{\l_1 - \l_2}{2\d} \ , \quad p = \frac{\l_1 - 2 \l_2 + \l_3}{2\d} \ .
\eea
We would like to have the distribution of $\d$, $e$ and $p$, on the condition that we are at an extremum of the density field. For $\d$, the probability distribution conditioned on the existence of an extremum was calculated in \cite{Bardeen:1986}. The unconditional distribution of $e$ and $p$ was worked out in \cite{Zeldovich:1970, Doroshkevich:1970}, and the conditional joint distribution for $e$ and $p$, given $\d$ (and assuming $\d>0$), is given in \cite{Sheth:1999su} (equation A3).
However, finding the distribution of $e$ and $p$ under the additional condition that the density field has an extremum is not straightforward. Note that $\l_i$ are the eigenvalues of the tidal tensor, and they determine whether the gravitational potential $\Phi$ has an extremum, and this does not necessarily coincide with an extremum of the density contrast $\d$. The tidal matrix and the matrix given by the Hessian of $\d$ do not commute, so they cannot be diagonalised simultaneously, and finding the distribution of the eigenvalues of the former under a condition for the eigenvalues of latter is a non-trivial problem \cite{Angrick:2013wba}.
For the number density of peaks/troughs we use the result of \cite{Angrick:2013wba} (equation 37). It requires as input the values of $e$ and $p$. For the distribution of $e$ and $p$ we use the result of \cite{Sheth:1999su}, modified to take into account that we can have $\d<0$. This distribution has been conditioned on the value of $\d$, but not on the requirement that there is a peak/trough. This is not expected to make a large difference (see \eg figure 3 of \cite{Angrick:2013wba}), because regions with large density contrast often contain extrema, and regions with small density contrast don't have a large effect. As the overall impact of deviations from spherical symmetry turns out in our case to be small, conditioning the distribution of $e$ and $p$ on the existence of an extremum would be a small correction to a small effect.

\subsection{The average expansion rate}

\para{Ensemble of ellipsoids.}

\begin{figure}[t]
  \centering
  \includegraphics[width=0.7\textwidth]{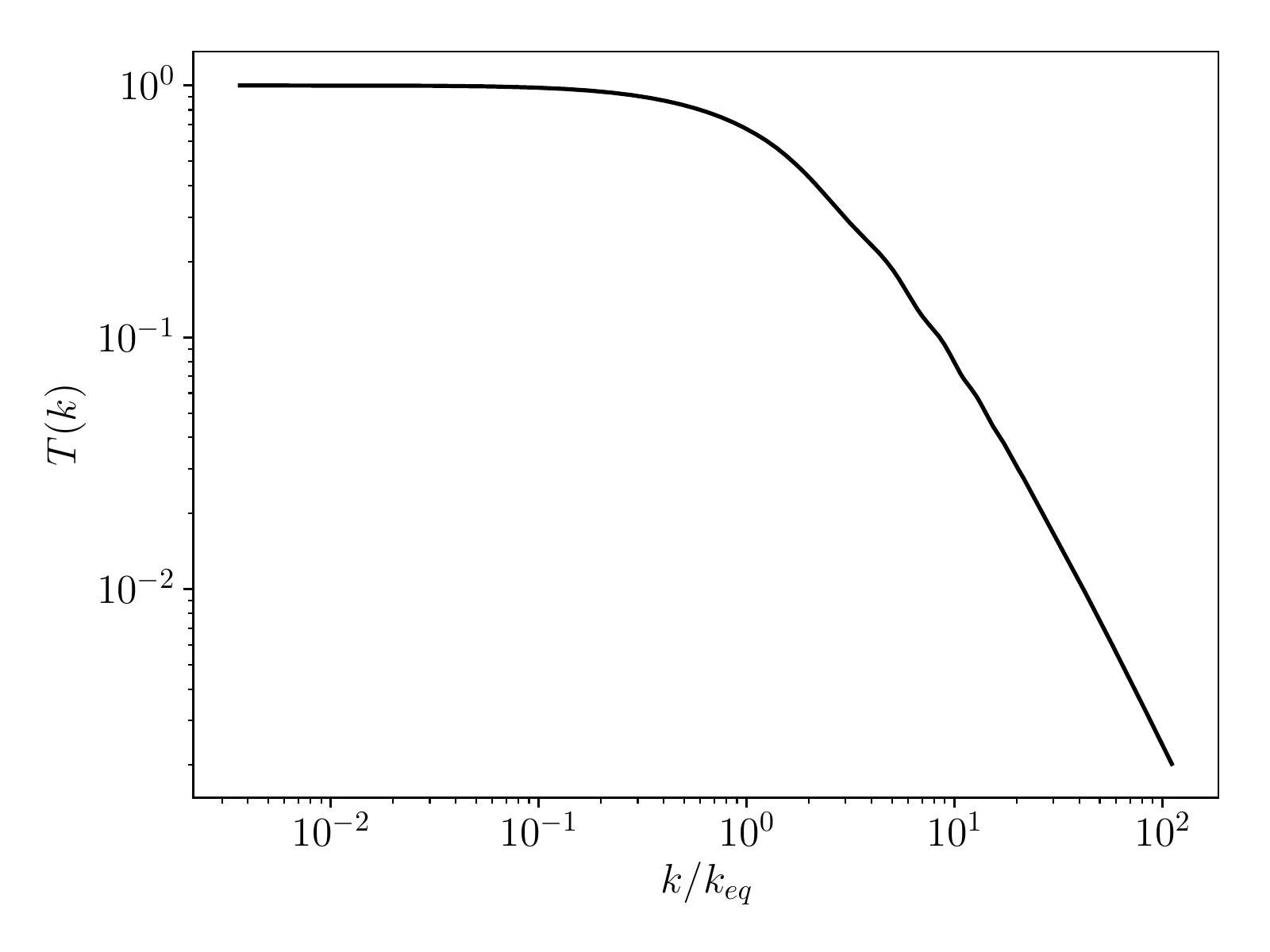}
  \caption{Transfer function of a cold dark matter (CDM) model with
    $\ob=0.02$ and $\om=0.14$ and no dark energy, calculated with CAMB
    \cite{Lewis:1999bs}. Wavenumbers are in units of the
    matter-radiation equality scale $k_{eq}$.}
  \label{fig:Tk}
\end{figure}

We consider a model where the volume of the universe is divided into peaks/troughs, which evolve according to the ellipsoidal model discussed above, and smooth regions, which evolve like the background. As the peaks/troughs expand and collapse and their volume fractions of change, the average expansion rate evolves. We calculate the average expansion rate as follows, generalising the spherical expression in \cite{Rasanen:2008it} to include ellipticity and prolateness:
\bea \label{statH}
  H(t) &=& \int\rmd\d\rmd e\rmd p \, v(\d,e,p,t) H(\d,e,p,t) \el
  &=& \frac{ \int\rmd\d\rmd e\rmd p \, f(\d,e,p,t) s(\d,e,p,t) H(\d,e,p,t) }{  \int\rmd\d\rmd e\rmd p \, f(\d,e,p,t) s(\d,e,p,t)} \ ,
\eea
where $v(\d,e,p,t)$ is the fraction of volume, at time $t$, in regions
that had initial linear density contrast $\d$, ellipticity $e$ and
prolateness $p$, and
$H(\d,e,p,t)=\frac{1}{3}\sum_i\frac{\adot_i}{a_i}$ is the expansion
rate, at time $t$, of a region with such initial conditions. The
volume fraction $v=s f/[\int\rmd\d\rmd e\rmd p \, s f]$ has two
parts. The factor $s(\d,e,p,t)\equiv a_1 a_2 a_3/{\bar a}^3$ is the
volume of a region with initial linear density contrast $\d$,
ellipticity $e$ and prolateness $p$, relative to the background
volume, at time $t$. It depends only on how the region expands
relative to the background.  The factor $f(\d,e,p,t)$ is the number
density of extrema with initial $\d,e$ and $p$, at time $t$. It
depends on how the number density of peaks and troughs changes with
time. The peak number density is defined for a given smoothing scale $R$ and thus depends on the choice of $R$; we use a Gaussian window function. We take the primordial power spectrum
to be a power law with spectral index $n=0.96$ (at the pivot scale $k=0.05$
Mpc$^{-1}$) \cite{Ade:2015xua}. We take the transfer function of a
cold dark matter (CDM) model with $\ob=0.02$ and $\om=0.14$ and no
dark energy, calculated with CAMB\footnote{\url{http://camb.info}} \cite{Lewis:1999bs} and plotted in
figure~\ref{fig:Tk}. These values, usually determined in the \LCDM
FRW model \cite{Ade:2015xua}, are model-independent
\cite{Vonlanthen:2010cd, *Audren:2012wb, *Audren:2013nwa}.

As in \cite{Rasanen:2008it}, we fix the smoothing scale $R(t)$ by demanding that the root mean square density contrast is unity at all times, $\sigma_0[R(t),t]=1$. The timescale of changes is thus determined by location of the turnover in the transfer function, which comes from the matter-radiation equality time $\teq=5\times10^4$ yr.
The idea is that $R$ determined like this gives the scale for typical structures. The precise numbers of our results depend on the choice of value for $\sigma_0(R,t)$, as we will see in \sec{sec:res}, but as in \cite{Rasanen:2008it}, the qualitative behaviour and the order of magnitude are the same as long as $\sigma_0(R,t)$ is of order unity. As density perturbations evolve in time, $R$ grows to compensate, modelling the merging of structures into larger ones. Volume that is neither in peaks nor troughs is taken to evolve like the background. We fix the normalisation of the number density by demanding that in the asymptotic future all volume will be in peaks or troughs.

In the same way, the average shear scalar is calculated as
\bea \label{statsigma}
  \av{\sigma^2} &=& \frac{1}{3} \int\rmd\d\rmd e\rmd p \, v(\d,e,p,t) \left( \sum_i \frac{\adot_i^2}{a_i^2} - \sum_{i<j} \frac{\adot_i}{a_i} \frac{\adot_j}{a_j} \right) \ .
\eea

\section{Results} \label{sec:res}

\subsection{The average expansion rate}

We calculate the averages \re{statH} and \re{statsigma} numerically. The functions $a_i$ are calculated from \re{aia} with the initial conditions \re{aiin}, using the initial distributions for $\l_i$ discussed in \sec{sec:ell}, with the virialisation condition based on the tensor virial theorem given in \cite{Angrick:2010qg}, and the non-linear prescription for the external tides. For comparison, we present the results for the case when the peaks are spherical, calculated with the same primordial power spectrum and transfer function.

\begin{figure}[t]
  \centering
  \includegraphics[width=0.7\textwidth]{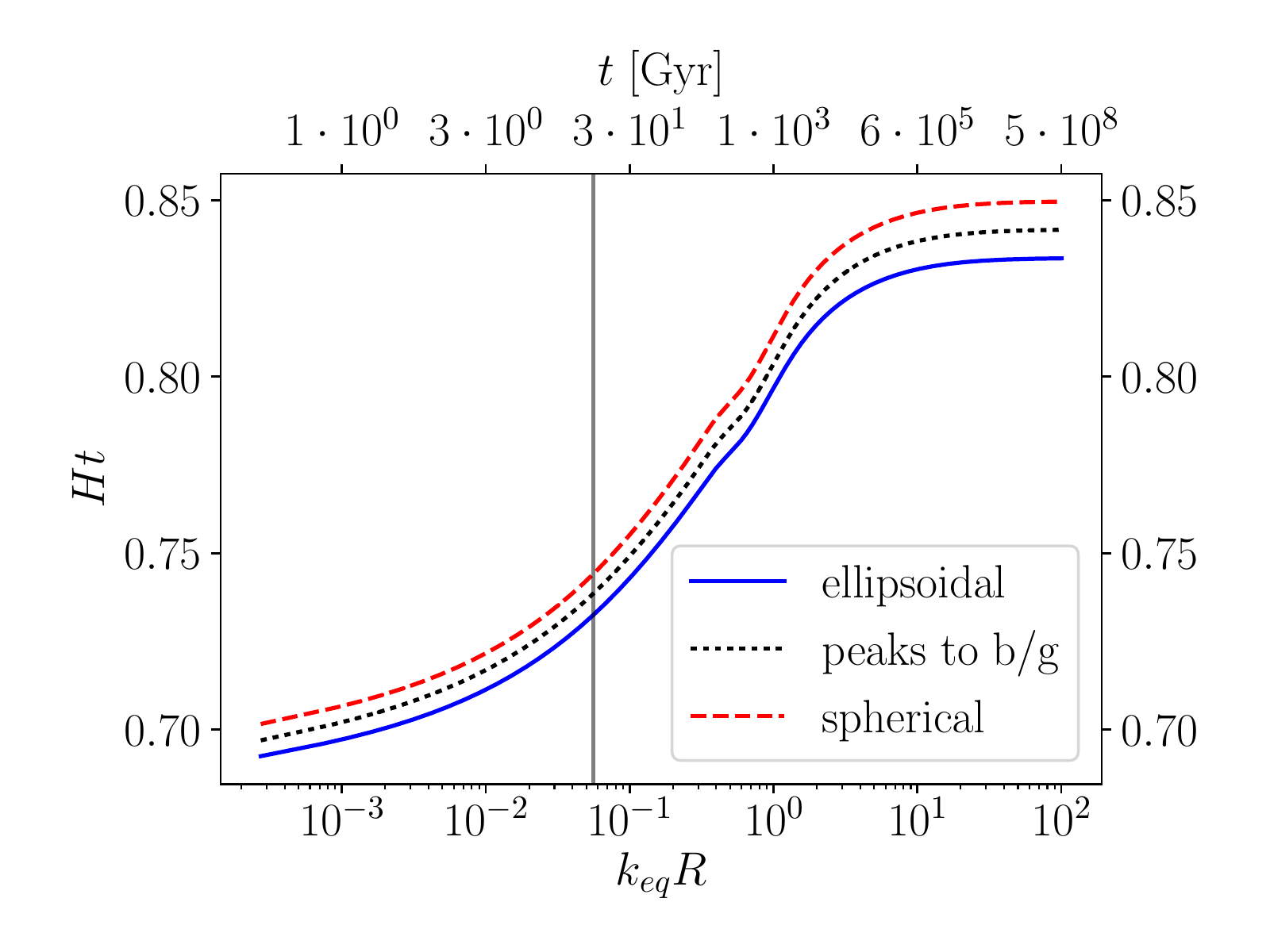} \\
  \caption{The expansion rate times the age of the universe, $Ht$, for the full ellipsoidal case (solid blue), spherical case (red dashed) and the ellipsoidal case when peak volume is assigned to the background (black dotted). The axis at the top shows the age of the universe corresponding to the $k_{eq} R$ value on the bottom axis. The vertical line corresponds to $t=14$ Gyr.}
  \label{fig:compHt}
\end{figure}

In \fig{fig:compHt} we show the expansion rate $Ht$ as a function of
$\keq R$ ($\keq$ is the comoving wavenumber of the modes that cross
the Hubble radius at matter-radiation equality) and $t$. At early
times, the expansion rate is close to the background Einstein--de
Sitter case $Ht=2/3$ (we do not consider radiation, so the model does
not apply at times earlier than $\sim10^6$ years). As the underdense
regions that expand faster than the background take up more of the
volume, $Ht$ rises, with a transition at $R\sim\keq^{-1}$,
corresponding to $t \approx 10^7\teq\approx10^3$ Gyr. The value practically saturates at $Ht=0.83$ at very late times $t\approx 10^5$ Gyr. (Note that the age at late times is exponentially sensitive to the value of $R$.)

Because the ellipsoids are homogeneous, the expansion is slower than
in the spherical case, as discussed in \sec{sec:Buchert}. Therefore
the expansion rate in the ellipsoidal case is always smaller than in
the spherical case, where the asymptotic value is $Ht=0.85$. (An
ellipsoid can expand faster than a spherical region in the sense that
it may continue expanding in one or two directions after a sphere has
already collapsed, but the effect of such regions on the average is
small.) The fact that every individual region expands slower does not
necessarily mean that the average expansion rate is smaller, nor that
the corresponding acceleration is smaller, because $\sQ$ is not
additive. In fact, having some regions slow down more can lead to
acceleration \cite{Kai:2006ws, Rasanen:2006zw, Rasanen:2006kp,
  Boehm:2013qqa}. Nevertheless, in our case the shear is too small and
its rise is not rapid enough to give acceleration.

\begin{figure}[t]
  \centering
  \includegraphics[width=0.7\textwidth]{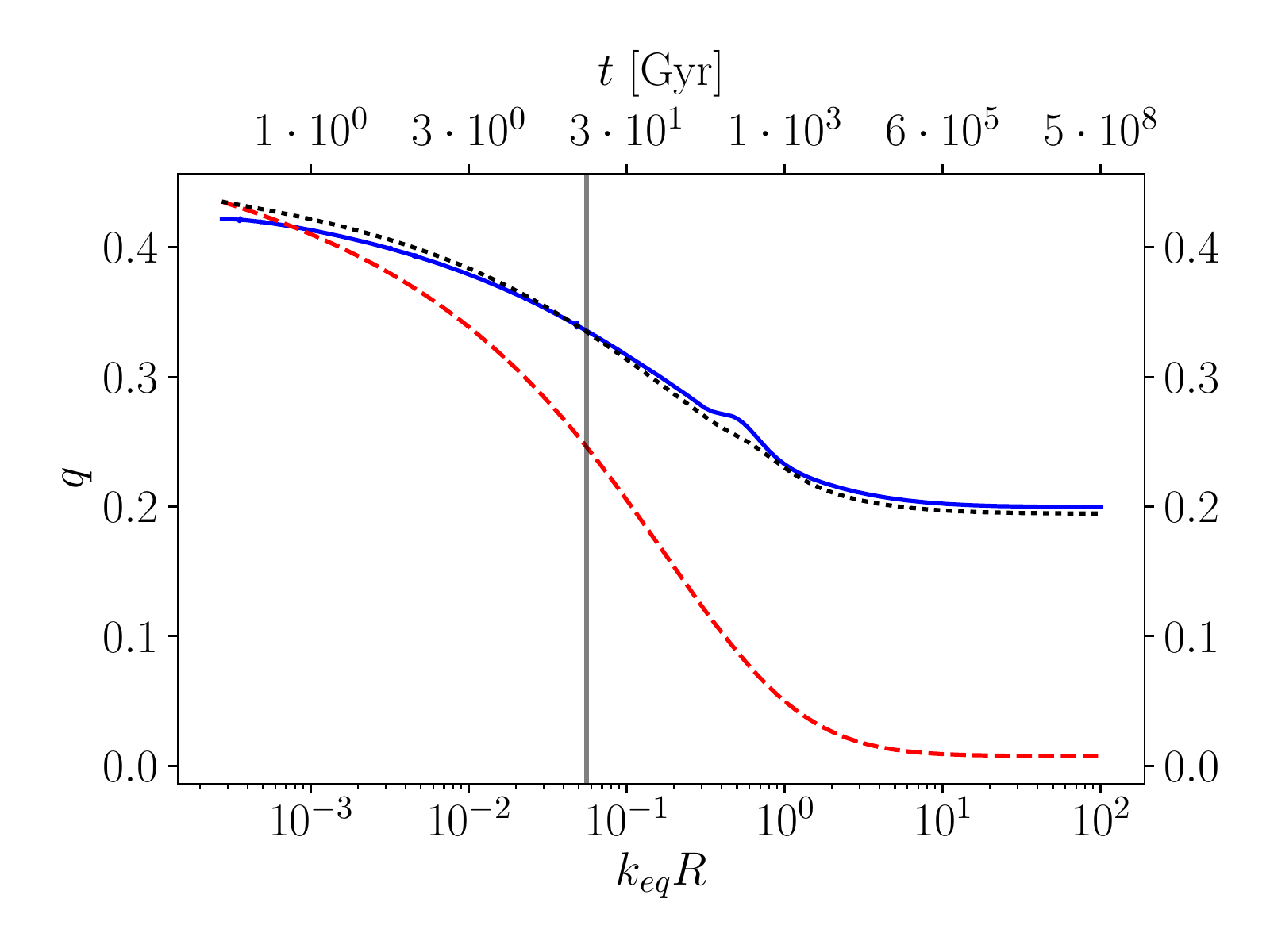} \\
  \caption{The deceleration parameter $q$ determined by calculating
    $Ht$ from the ensemble and using \re{q} (blue), calculating $Ht$
    and the proper volume from the ensemble to give $\Om$ and using
    \re{q} (black dotted) or calculating $Ht$ from the ensemble and
    finding $\Om$ from $Ht$ (red dashed). See \sec{sec:disc} for
    details.}
  \label{fig:q}
\end{figure}

\begin{figure}[t]
  \centering
  \includegraphics[width=0.7\textwidth]{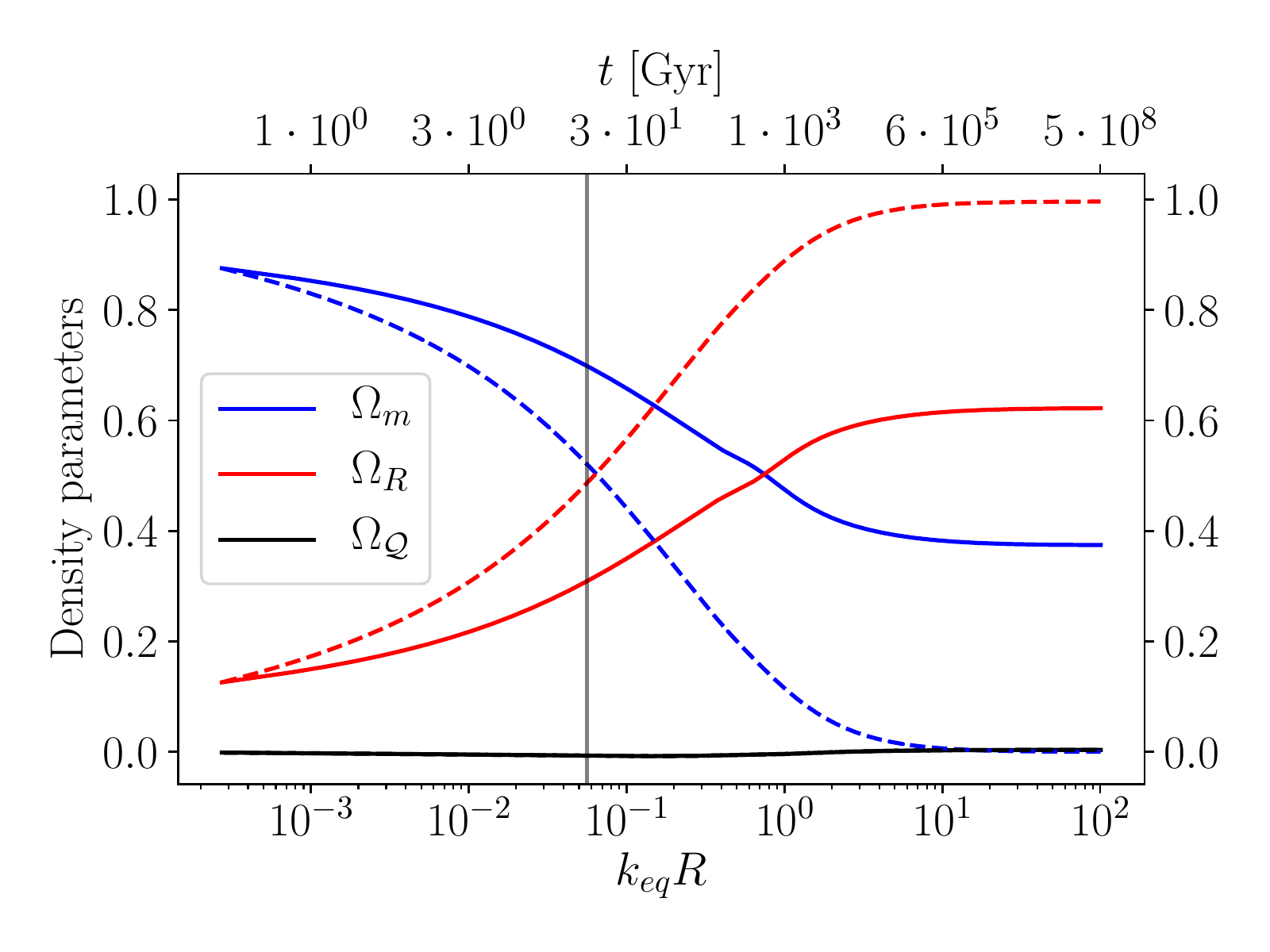} \\
  \caption{The density parameters $\Om$ (blue), $\OR$ (red) and $\OQ$
    (black). Solid lines correspond to $\Om$ determined using \re{q},
    dashed lines to the case when $\Om$ is calculated from $Ht$. See
    \sec{sec:disc} for details.}
  \label{fig:omegas2}
\end{figure}

The deceleration parameter $q$ is plotted in \fig{fig:q}. The blue
line is obtained using the first equality in \re{q},
$q=-\Hdot/H^2-1$. We always have $q>0$, so the expansion decelerates;
at late times, the expansion rate asymptotes to $Ht=0.83$, which
corresponds to $q=0.2$. We can then determine $\Om$ by the second
equality in \re{q} as $\Om=2q-4\OQ$, and the spatial curvature density
parameter from \re{omegas} as $\OR=1-\OQ-\Om$. The resulting density
parameters are plotted in \fig{fig:omegas2} (solid lines). The
universe transitions around $R\sim\keq$ from being close to
matter-dominated to the asymptotic values $\Om=0.37$, $\OR=0.62$. The
backreaction variable remains small throughout, $|\OQ|<0.01$.
\begin{figure}[t]
  \centering
  \includegraphics[width=0.7\textwidth]{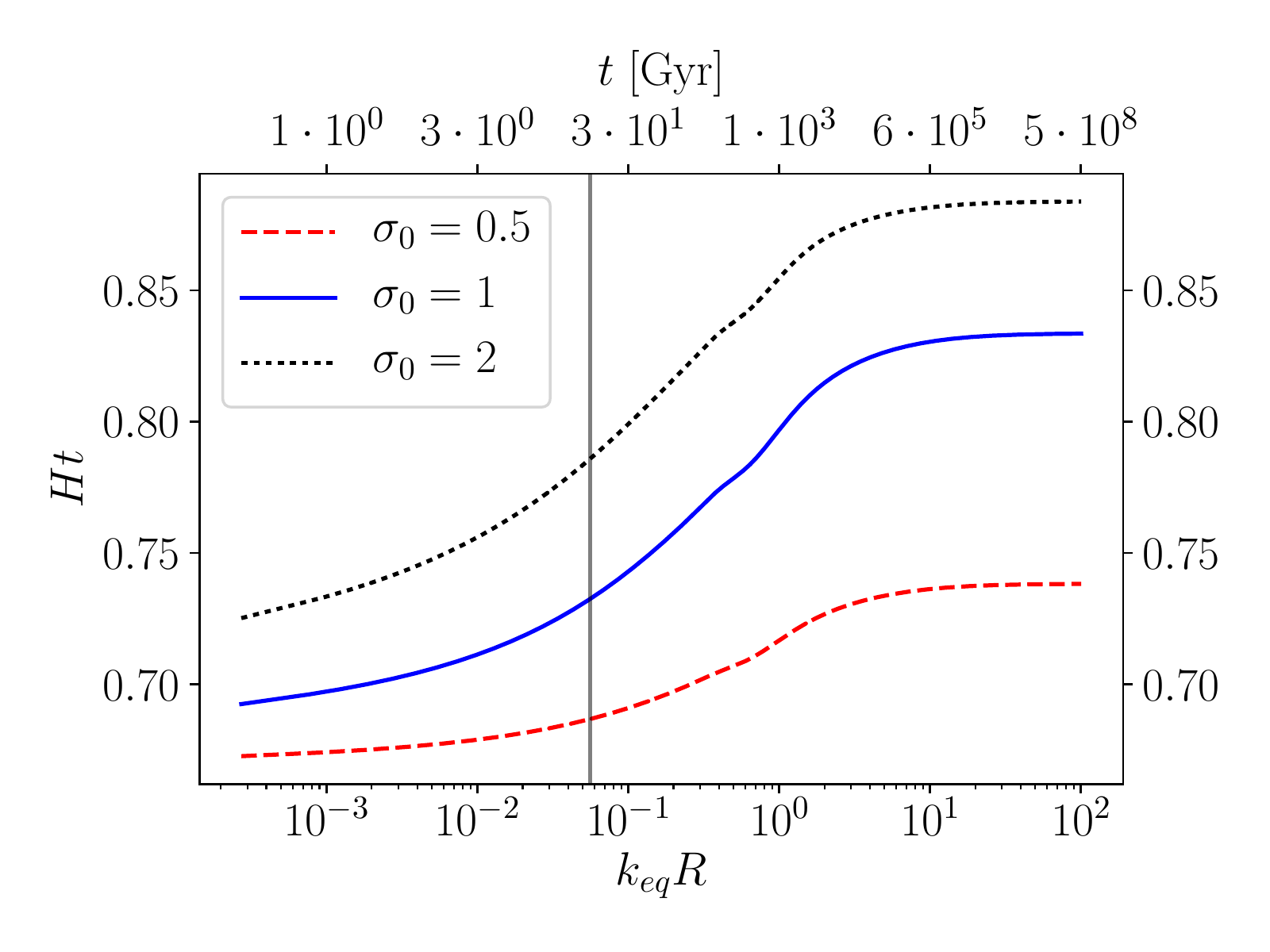} \\
  \caption{The expansion rate $Ht$ for $\sigma_0(R,t)$ equal to 0.5
    (red, bottom), 1 (blue, middle) and 2 (black, top). Note that the top $t$ values correspond to the case $\sigma_0(R,t)=1$. For the other cases they are larger or smaller by a factor of $2^{3/2}\approx3$.}
  \label{fig:difRHt}
\end{figure}

The precise numbers depend on how the smoothing scale has been set. In
\fig{fig:difRHt} we show $Ht$ in the cases when $\sigma_0(R,t)$ is
chosen to be 0.5, 1 or 2. A factor of 2 variation in the linear
density contrast around unity typically corresponds to a large change
in the non-linear evolution; in the spherical model, regions with
linear density contrast of 2 have already completely collapsed. We see
that the magnitude of the change in $Ht$ relative to the Einstein-de
Sitter value 2/3 goes from 11\% to 25\% and 32\% for the cases
$\sigma_0(R,t)=$ 0.5, 1 and 2, respectively. However, the qualitative
behaviour and the order of magnitude of the change and its timing stay
the same.

\begin{figure}[t]
  \centering
  \includegraphics[width=0.7\textwidth]{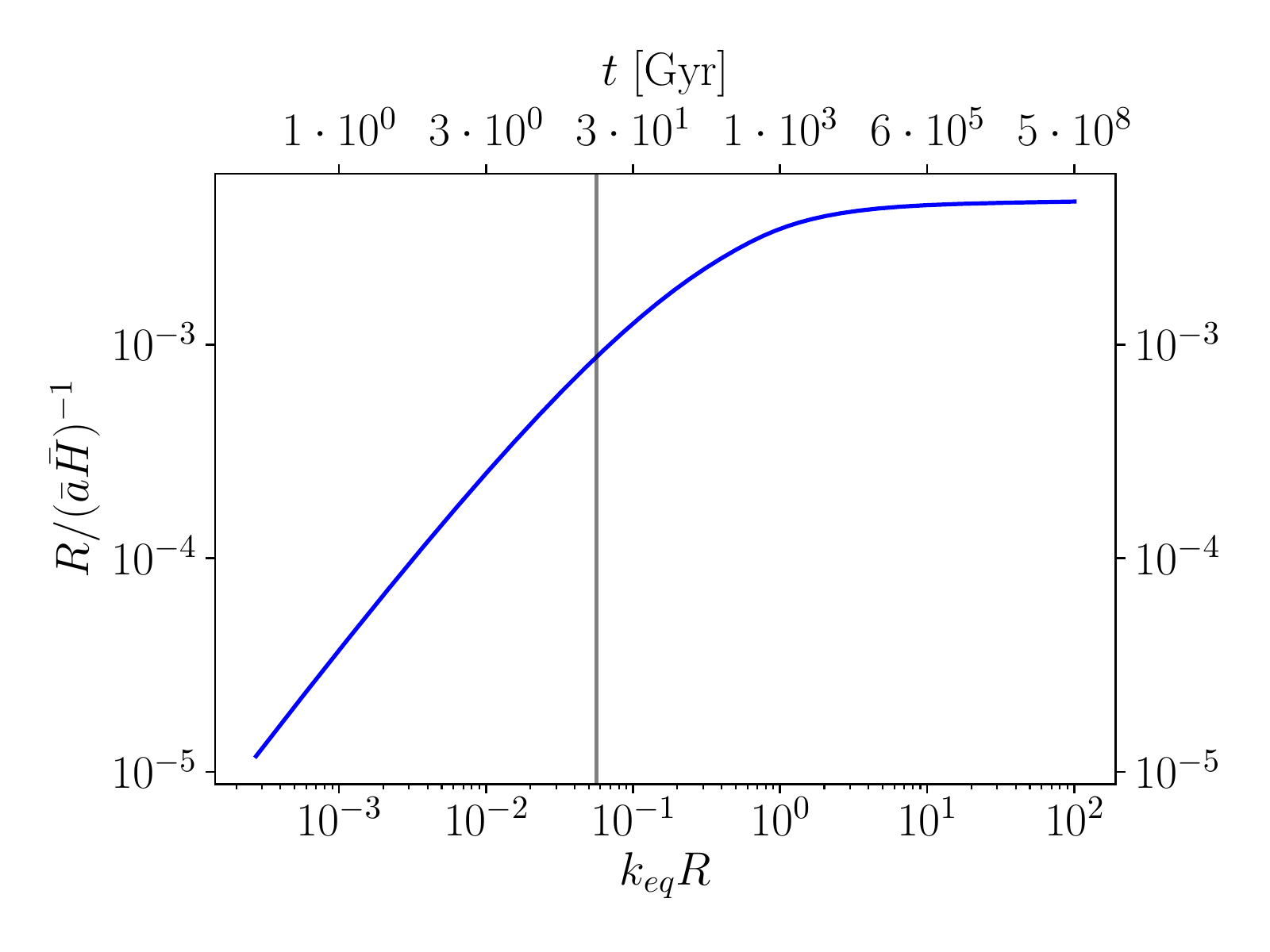} \\
  \caption{The smoothing scale, corresponding to the typical size of
    structures, relative to the background Hubble length,
    $R/(\bar a \bar H)^{-1}$.}
  \label{fig:scale}
\end{figure}

The transition era $R\sim\keq^{-1}$ is also visible in
\fig{fig:scale}, where we compare the smoothing scale $R(t)$
(indicative of the size of largest structures at time $t$) to the
comoving background Hubble length $(\bar a \bar H)^{-1}$. Structures
start small, their relative size grows rapidly and continues to grow
slowly after the transition. The reason is that for $k<\keq$, the transfer function is essentially constant. If the
primordial spectrum were scale-invariant, then there would be no scale
in the system any more, and dimensionless quantities like
$R/(\bar a \bar H)^{-1}$ would be constant. With a non-scale-invariant
spectrum, $n=0.96$, there is still slow evolution.

\subsection{The shear}

\begin{figure}[t]
  \centering
  \includegraphics[width=0.7\textwidth]{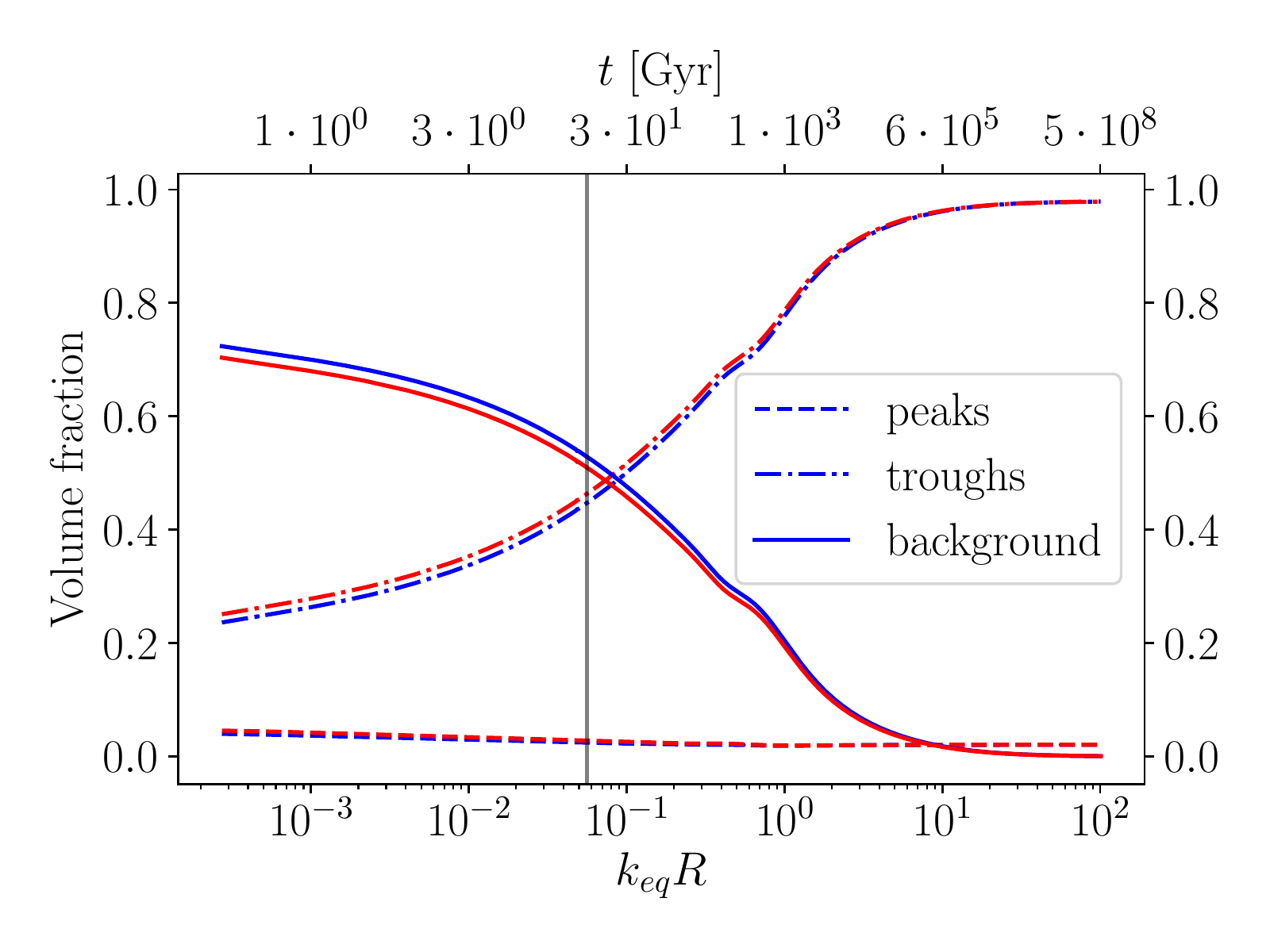} \\
  \caption{The fraction of volume in peaks (dashed), in troughs
    (dot-dashed) and in the background (solid). The blue line is the
    ellipsoidal case and the red line is the spherical case.}
  \label{fig:volfrac}
\end{figure}

\begin{figure}[t]
  \centering
  \includegraphics[width=0.7\textwidth]{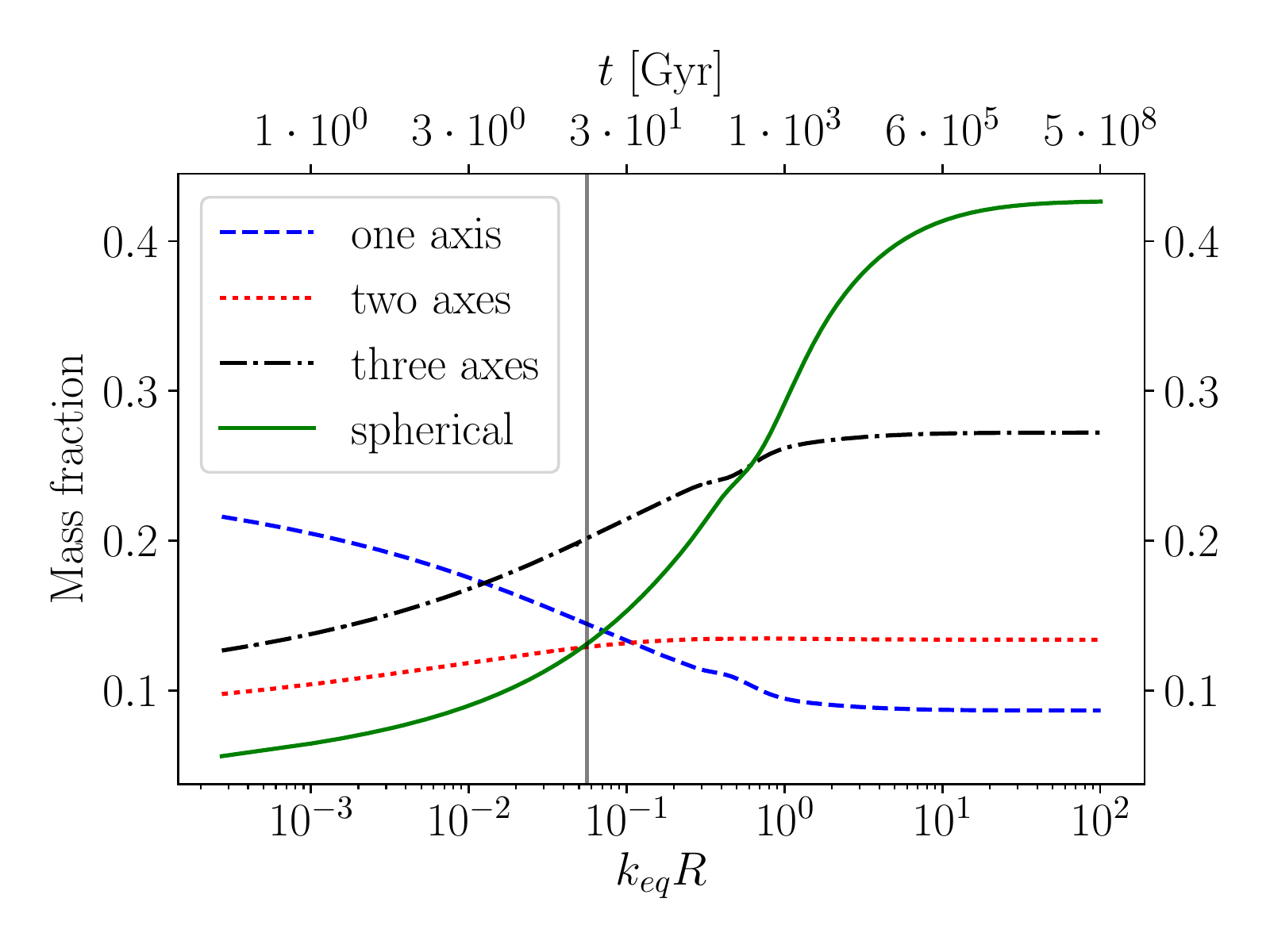} \\
  \caption{The fraction of regions that have collapsed along 1, 2 or 3
    axes (not weighted by volume). The spherical case is shown for comparison.}
  \label{fig:collfrac}
\end{figure}

In \fig{fig:volfrac} we show the fraction of volume in the peaks, troughs and the background. The results are almost the same in ellipsoidal and spherical cases. Between $\keq R=0.1$ and $\keq R=1$ (corresponding to $t=30$ Gyr and $t=10^3$ Gyr, respectively) troughs start to dominate the volume. This happens both because their fraction of initial volume (which is also the mass fraction) grows due to merging and because they expand faster than the background. In \fig{fig:collfrac} we show the mass fraction of regions that have collapsed along one, two or three axes, corresponding to walls, filaments and clusters.  The transition is clear also here, and the fractions practically saturate around $R=\keq^{-1}$. (After the transition, the volume fraction of all these regions is negligible, as \fig{fig:volfrac} shows.)  Unlike the spherical model, the ellipsoidal case allows underdense regions to collapse due to decelerating effect of shear, but the troughs are so deep that this has negligible effect on the averages. The mass fraction of structures that have collapsed at least on one axis peaks at $49\%$. After the transition, $27\%$ of the mass is in clusters, $14\%$ in filaments and $8\%$ in walls. Compared to simulations and observations of structure, our model is missing efficient flow of mass from underdense to overdense regions \cite{Bond:1995yt, Cautun:2014fwa}.
\begin{figure}[t]
  \centering
  \begin{subfigure}[b]{0.48\textwidth}
    \includegraphics[width=\textwidth]{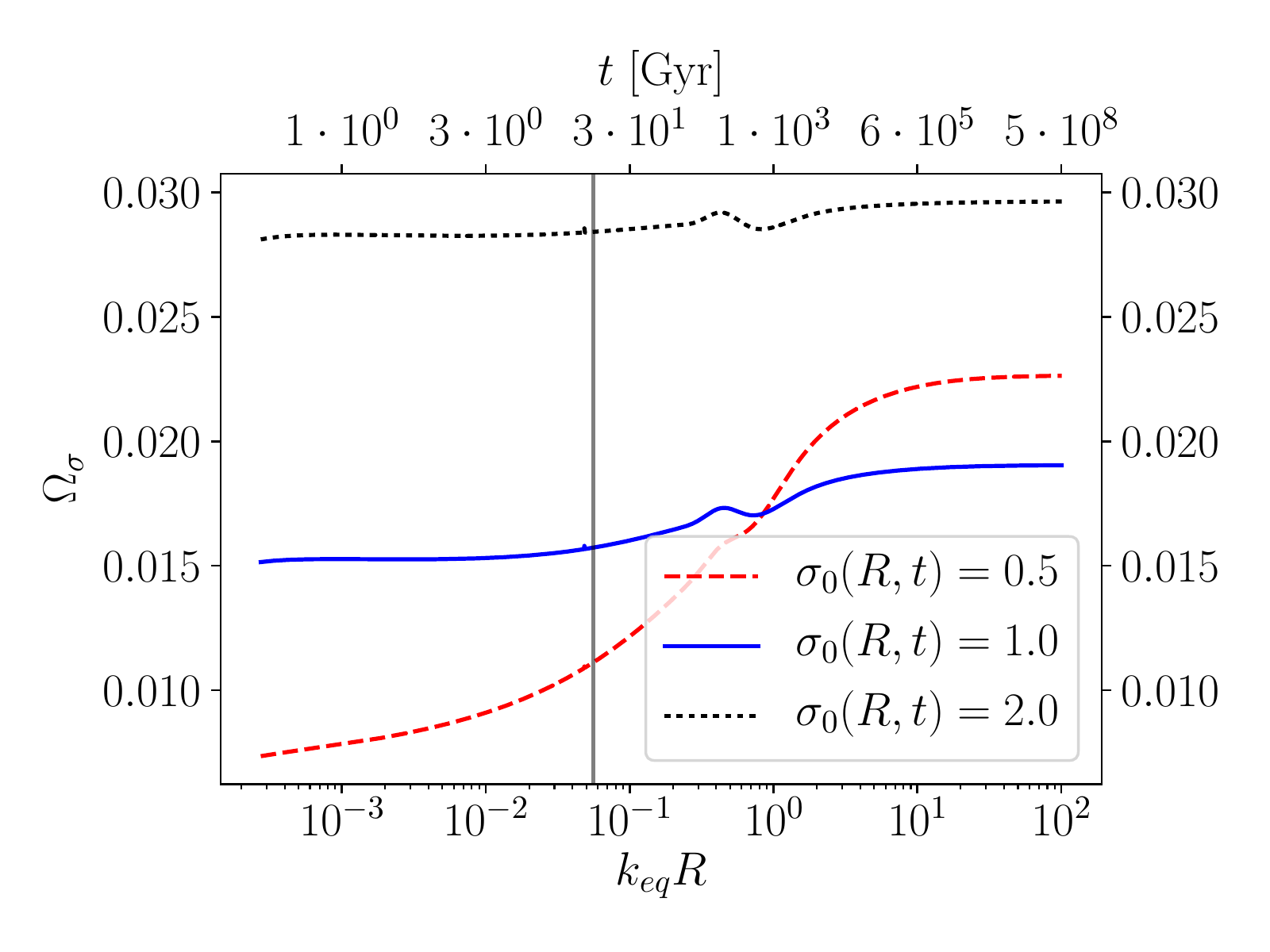}
    \caption{}
    \label{fig:sigma}
  \end{subfigure}
  ~
  \begin{subfigure}[b]{0.48\textwidth}
  \includegraphics[width=\textwidth]{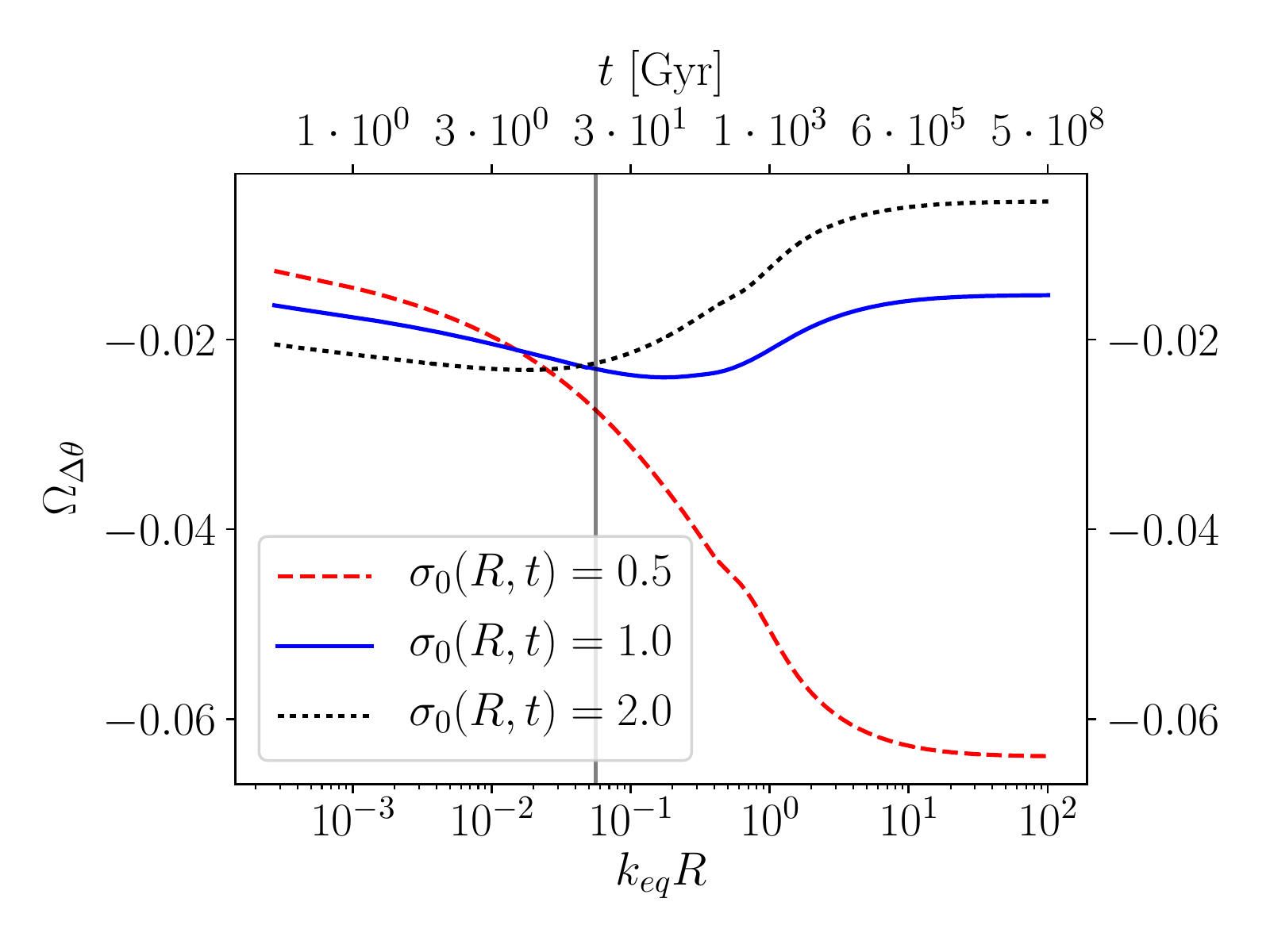}
  \caption{}
  \label{fig:difRvar}
  \end{subfigure}
  \\

  \caption{The shear density parameter $\Omega_{\sigma}$ (left) and
    the variance density parameter $\Ov$ (right) for $\sigma_0(R,t)$
    equal to 0.5 (red), 1 (blue) and 2 (black).}
  \label{fig:sigma_difRvar}

\end{figure}

Figure \ref{fig:sigma} shows the effect of the shear in terms of $\Omega_\sigma$ for different choices of $\sigma_0(R,t)$. For $\sigma_0(R,t)=1$, the amplitude is between 1\% and 2\%, an order of magnitude below what would be required to have a significant impact on the average expansion rate. In this case, unlike for the other quantities, the difference between the non-linear and hybrid treatments would be discernible by eye in the plots, but the absolute difference is only at the level 0.003. The variation due to changing the value of $\sigma_0(R,t)$ is much larger than this. If we set $\sigma_0(R,t)=2$, we keep more non-linear regions, resulting in larger shear. For $\sigma_0(R,t)=0.5$, there are fewer non-linear regions and the shear is initially smaller, but grows slightly larger at late times. These variations do not change the order of magnitude of the shear contribution, which remains below 3\%.

\begin{figure}[t]
  \centering
  \includegraphics[width=0.7\textwidth]{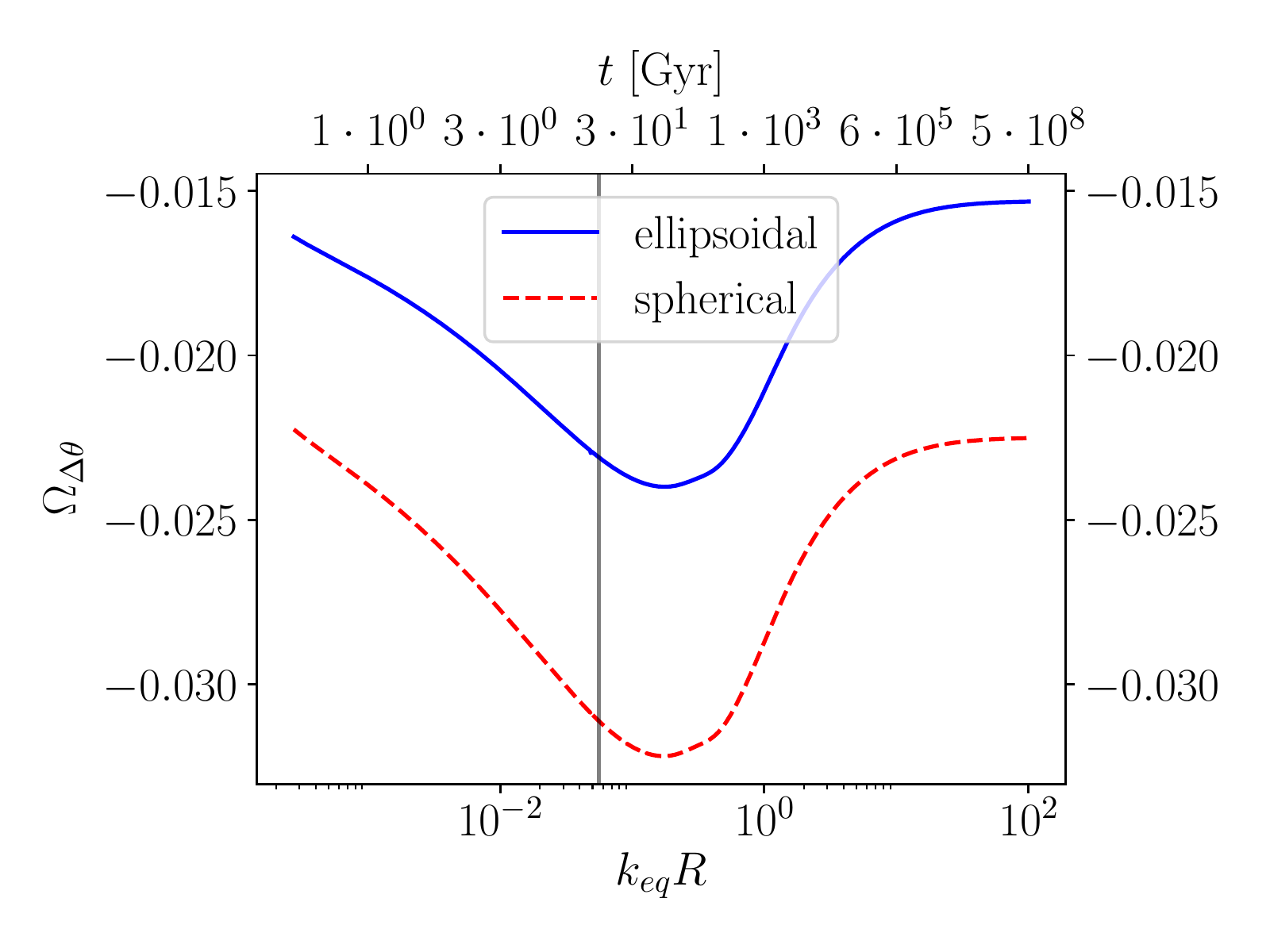} \\
  \caption{Comparison of the variance density parameter $\Ov$ for the spherical
    (red, lower) and ellipsoidal (blue, upper) cases.}
  \label{fig:var}
\end{figure}

The evolution of the variance is shown in \fig{fig:difRvar}. For $\sigma_0(R,t)=1$, at early times the variance is small because most of the volume evolves like the background, and at late times the variance is small because the volume is dominated by underdense regions with expansion rates close to each other (though far from the background). The transition era with peak variance is a bit before $R=\keq^{-1}$. In the ellipsoidal case the variance is smaller than in the spherical case by a factor of $1.5$, as shown in \fig{fig:var}, because the shear decelerates the expansion more, so troughs remain closer to the background.
As in the case of the shear, changing the value $\sigma_0(R,t)$ affects the qualitative evolution of $\Ov$, but not the order of magnitude, the maximum amplitude remaining smaller than 7\% in all cases.

\begin{figure}[t]
  \centering

  \begin{subfigure}[b]{0.48\textwidth}
    \includegraphics[width=\textwidth]{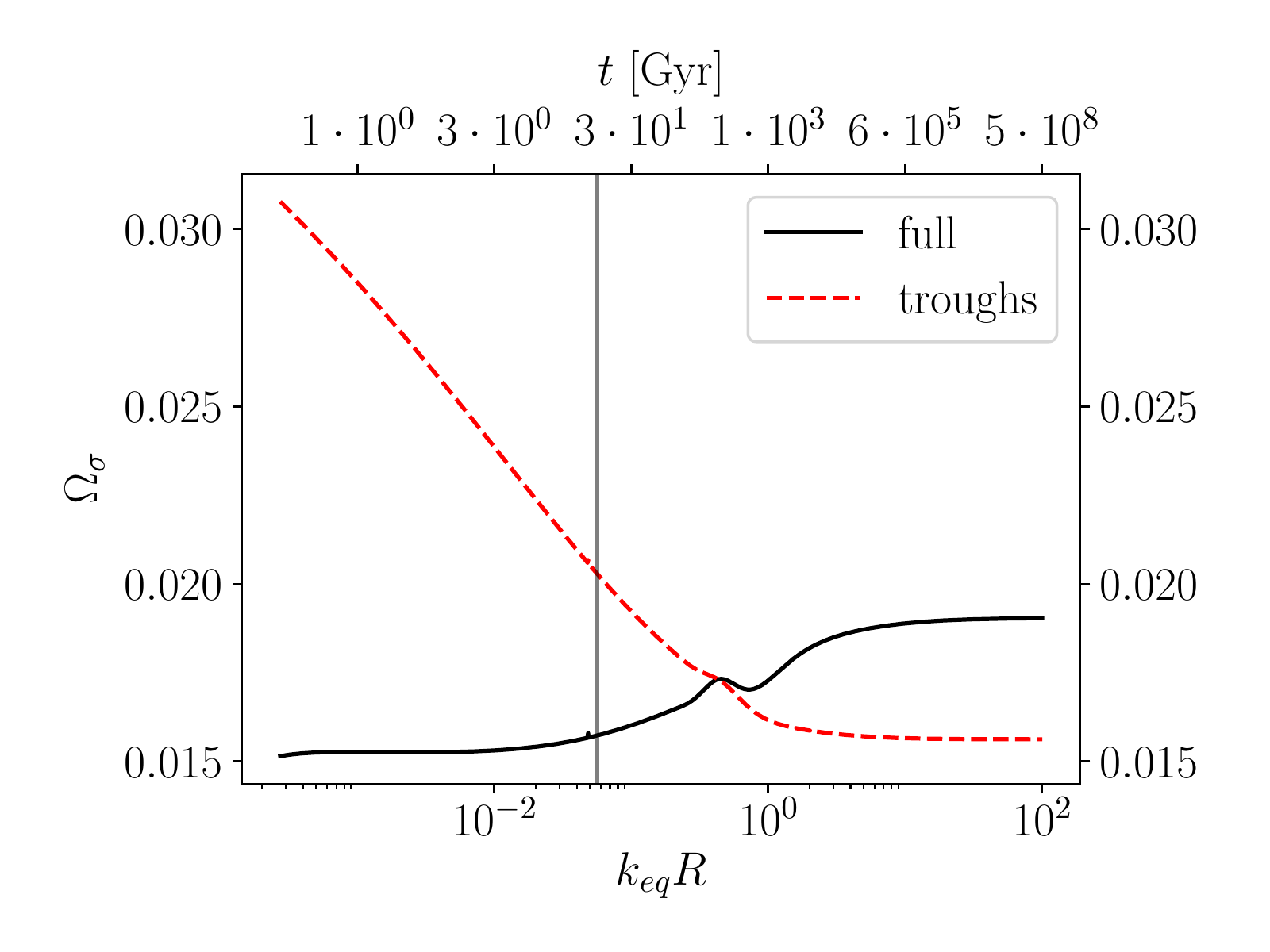}
    \caption{}
    \label{fig:troughsigma}
  \end{subfigure}
  ~
  \begin{subfigure}[b]{0.48\textwidth}
    \includegraphics[width=\textwidth]{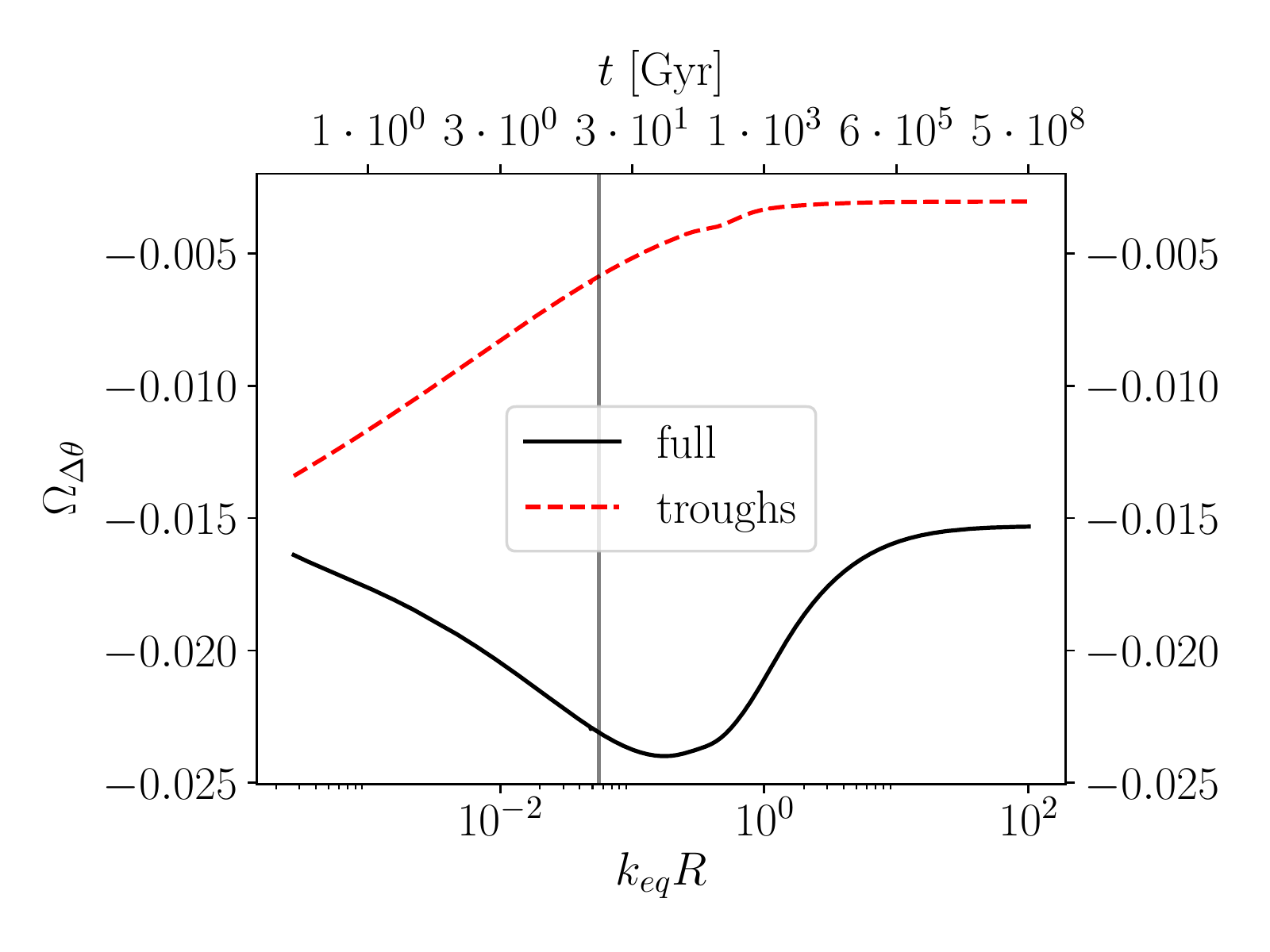}
    \caption{}
    \label{fig:troughvar}
  \end{subfigure}
  \\

  \caption{The shear density parameter $\Omega_\sigma$ (left) and the
    variance density parameter $\Ov$ (right) for the full result
    (black) and the troughs alone (red).}
  \label{fig:troughsigma_troughvar}
\end{figure}

In \fig{fig:troughsigma}, we show the shear density parameter $\Omega_\sigma$ for the troughs only, compared to the full result. The shear of troughs goes down with time, because they become more spherical as they expand. In contrast, the shear of peaks rises as the axes become more differentiated. The shear has a bump around $\keq$, where the peak number density grows rapidly, and peak contribution rises, followed by a dip as the trough contribution falls.
Figure \ref{fig:troughvar} shows $\Ov$ for the troughs and for the full result. Complementing the shear plot, it shows that the variance of the troughs goes down with time, but the overall variance amplitude has a bump around $\keq$ due to peaks becoming more differentiated, and their contribution to the average then goes down due to their shrinking volume.

\begin{figure}[t]
  \centering

  \begin{subfigure}[b]{0.48\textwidth}
    \includegraphics[width=\textwidth]{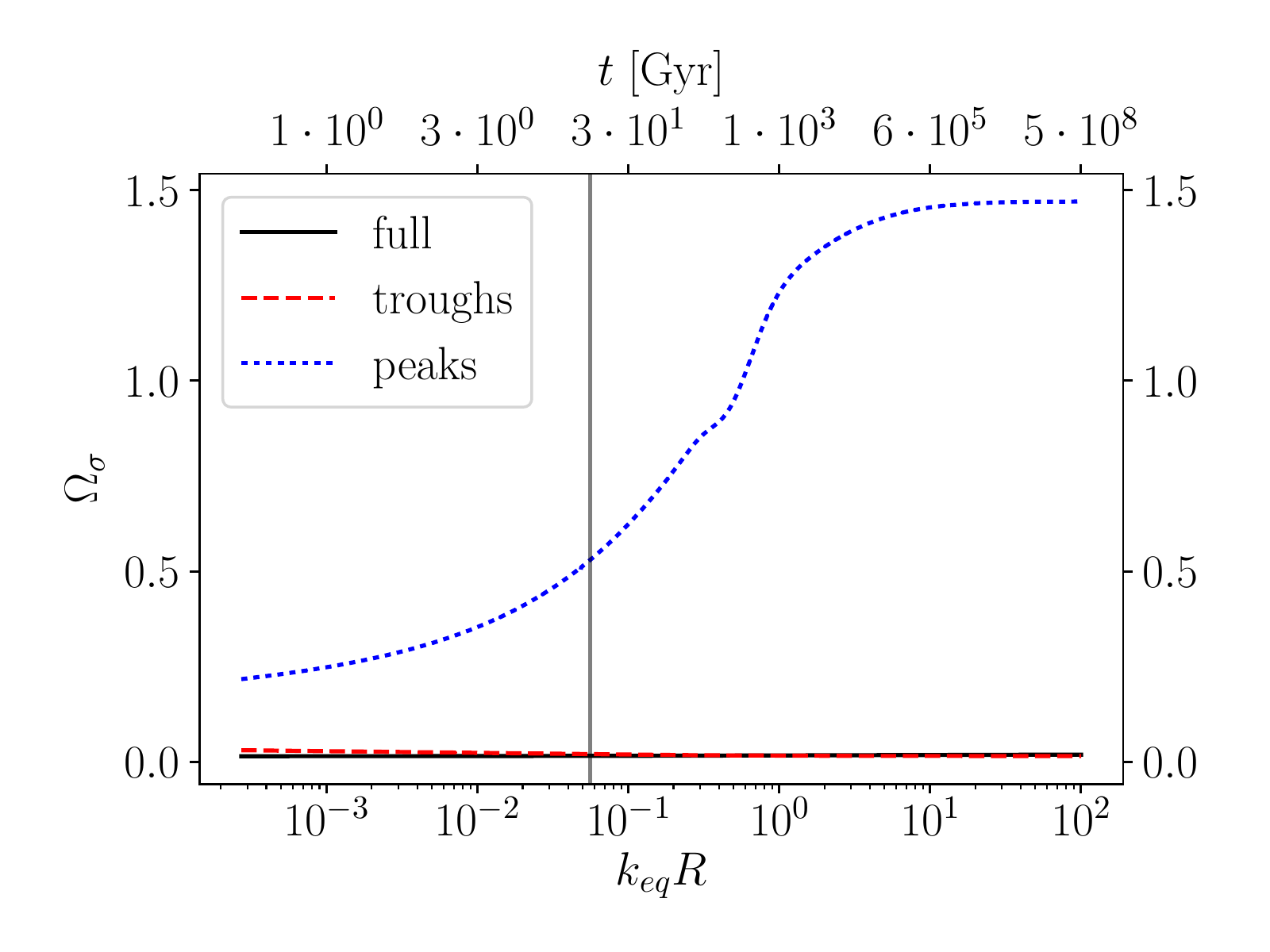}
    \caption{}
    \label{fig:troughpeaksig}
  \end{subfigure}
  ~
  \begin{subfigure}[b]{0.48\textwidth}
    \includegraphics[width=\textwidth]{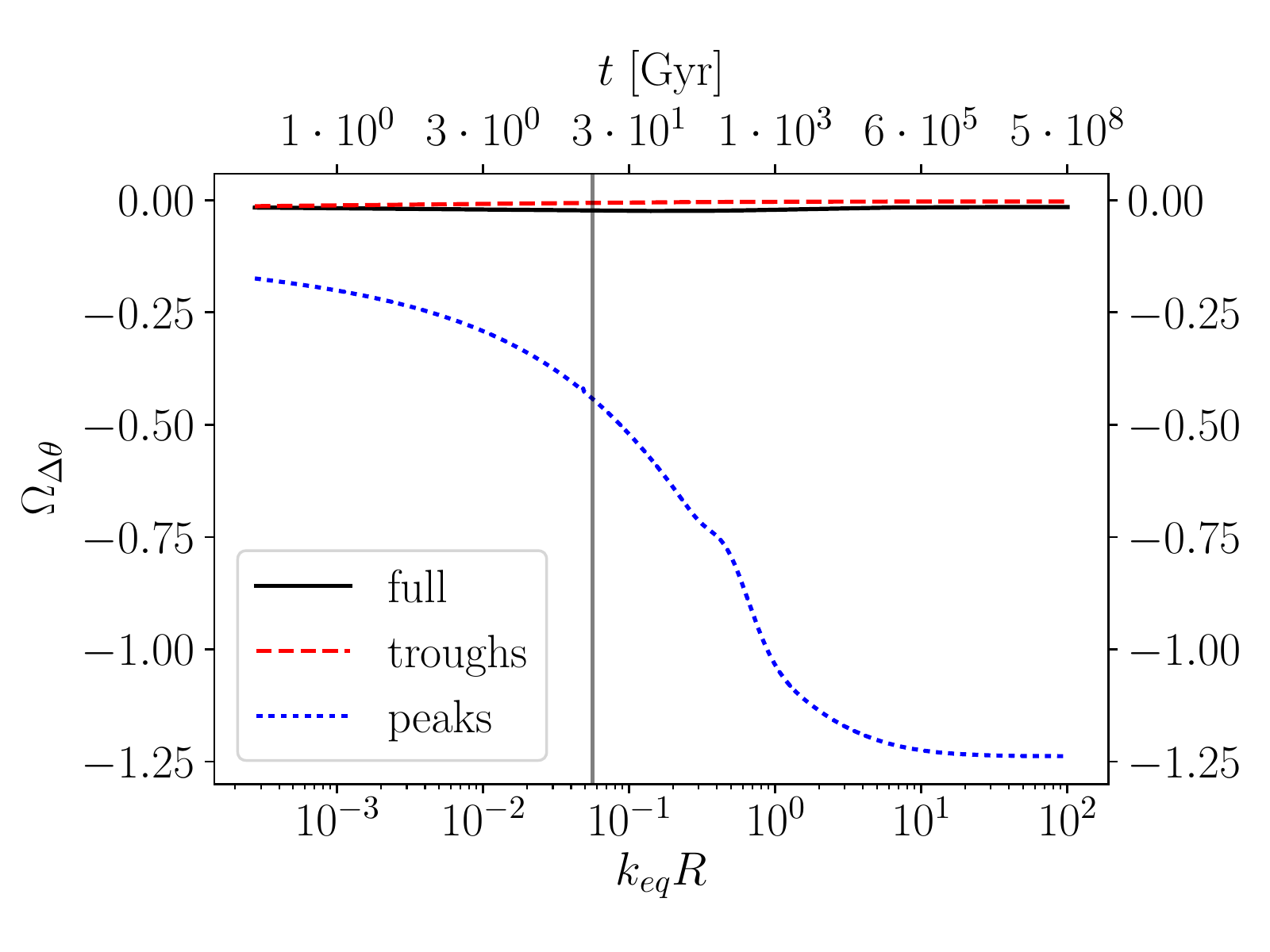}
    \caption{}
    \label{fig:troughpeakvar}
  \end{subfigure}
  \\

  \caption{The shear density parameter $\Omega_\sigma$ (left) and the
    variance density parameter $\Ov$ (right) for the full result
    (black), troughs (red) and peaks (blue).}
  \label{fig:troughpeaksig_troughpeakvar}
\end{figure}

\begin{figure}[t]
  \centering
  \includegraphics[width=0.7\textwidth]{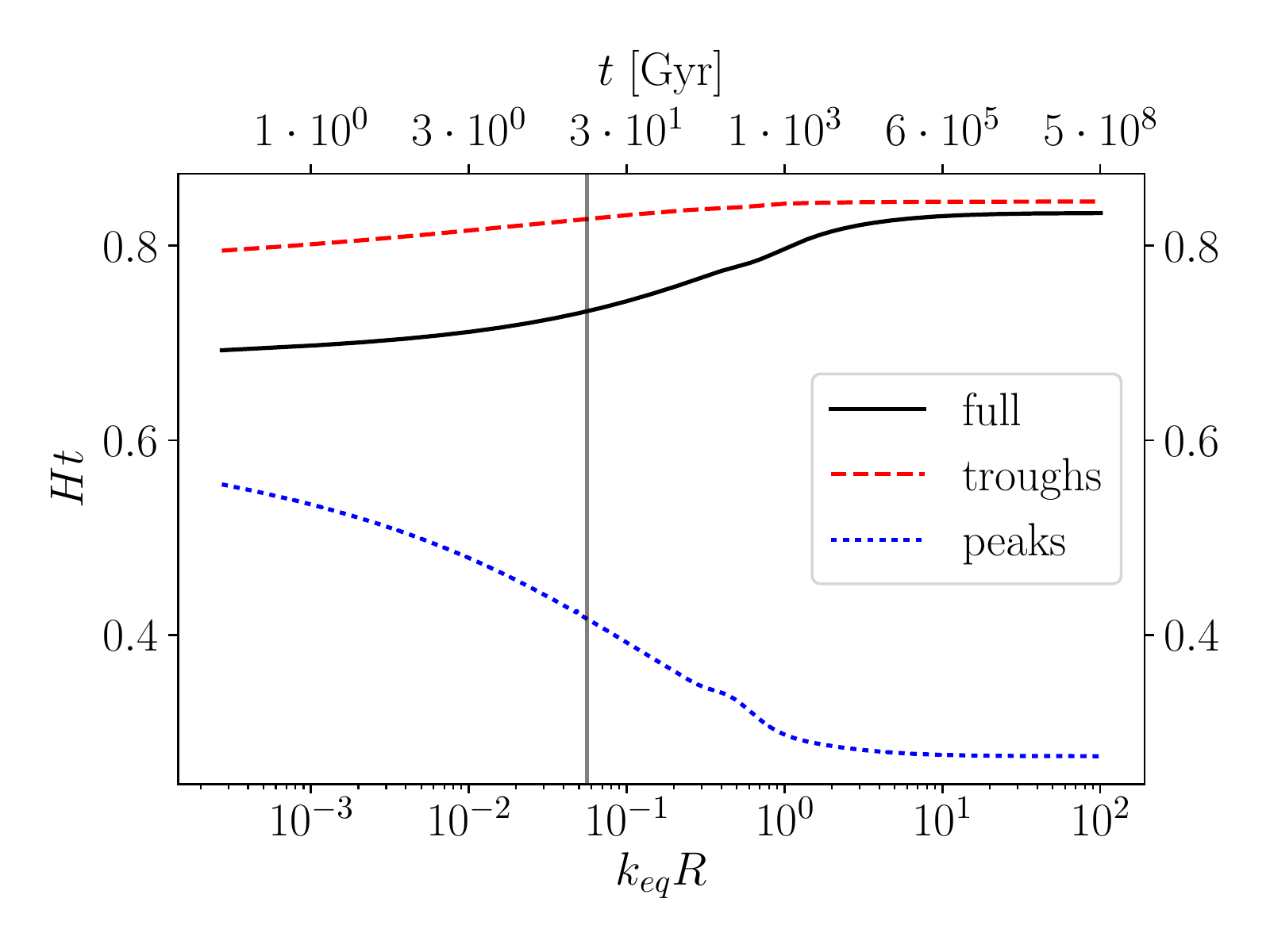} \\
  \caption{The expansion rate $Ht$ for the full result (black),
    troughs (red) and peaks (blue).}
  \label{fig:troughpeakHt}
\end{figure}

Figure \ref{fig:troughpeaksig} shows that the shear density parameter for the peaks alone grows large around $k=\keq$, to $\Omega_\sigma=1.5$. The evolution of the variance, shown in \fig{fig:troughpeakvar}, matches that of the shear, with $\Ov$ reaching the value $-1.2$ for the peaks. However, as the volume fraction of peaks simultaneously falls, their impact on the average expansion rate becomes smaller. In \fig{fig:troughpeakHt} we show $Ht$ for the peaks, troughs and the full sample of regions, which shows how the peaks bring down the expansion rate. In \fig{fig:compHt} the solid black line shows what happens if we allocate the volume of the peaks for the background (\ie their volume factor is as small as in the usual case, but they expand like the background).
This shows that the growth of $Ht$ can mostly be understood simply in terms of the faster-expanding troughs taking up more a larger fraction of space, but there is also a small contribution from the peaks, bringing $Ht$ down.

\begin{figure}[t]
  \centering
  \includegraphics[width=0.7\textwidth]{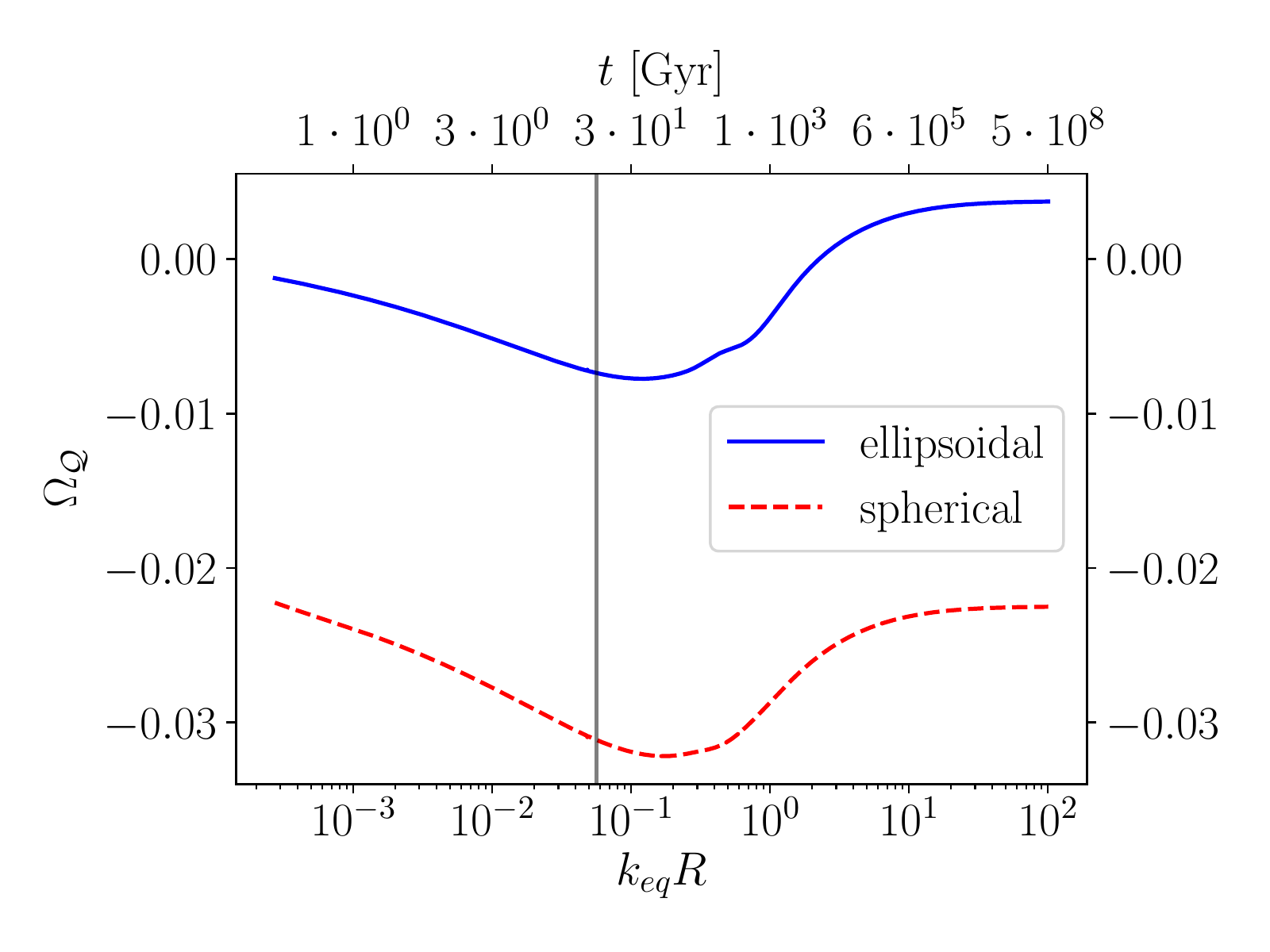} \\
  \caption{The backreaction density parameter $\OQ$ for the
    ellipsoidal (blue) and spherical (red) case.}
  \label{fig:OQ}
\end{figure}

The variance and shear contribute with the opposite sign to $\sQ$, so their contributions cancel to make $\OQ$, shown in figure \ref{fig:OQ}, smaller than either $\Omega_\sigma$ or $|\Ov|$ and negative, with an extremum amplitude of $\OQ=-0.008$ reached at $0.1\keq$. Such a close cancellation is not a robust result: as \fig{fig:sigma_difRvar} shows, for $\sigma_0=0.5$ the extremum value is $\OQ=-0.04$ and for $\sigma_0=2$ it is $\OQ=0.024$.
In the spherical case the variance and shear of each region cancel individually, and in the average over regions only the variance term appears, with an extremum of $\OQ=-0.03$. According to \re{q}, the observed values $q\approx-0.5$ and $\Om\approx0.3$ correspond to $\OQ\approx-0.3$, an order of magnitude above the largest numbers here.\footnote{The value $q\approx-0.5$ is the \LCDM model result. The value depends on the chosen parametrisation \cite{Shapiro:2005nz, *Gong:2006tx, *Elgaroy:2006tp, *Seikel:2007pk, *Seikel:2008ms, *Mortsell:2008yu, *Guimaraes:2009mp, *Serra:2009yp, *Cai:2010qp, *Wang:2010vj, *Park:2010xw, *Pan:2010zh, *Cai:2011px, *Li:2011wb, *Shafieloo:2012ht, *Aviles:2012ay, *Magana:2014voa, *Qing-Guo:2016ykt, *Ade:2015rim, Cattoen:2007id, *Visser:2009zs}. The value has also been derived under the assumption that light propagation is related to the expansion rate via the FRW distance-redshift relation, which is not necessarily the case if backreaction is large \cite{Rasanen:2008be, Rasanen:2009uw, Clarkson:2011br, Bull:2012zx, Boehm:2013qqa, Lavinto:2013exa}.}

\section{Discussion} \label{sec:disc}

\para{Successes and shortcomings.}

Structures have a large impact on the expansion rate, and $Ht$ grows by 25\% relative to the Einstein--de Sitter value $2/3$. This change of the same order of magnitude, but slightly smaller, than in the spherical case. The matter density parameter $\Om$ correspondingly falls by a factor of about 3 from 1 to $0.37$. These are asymptotic values, reached at $R= 9 k_{eq}^{-1}$ (corresponding to $t\sim10^5$ Gyr). The order of magnitude of the change from the Einstein--de Sitter case corresponds to present-day observations.\footnote{Note that the large value of the spatial curvature parameter $\OR=0.62$ is not in obvious contradiction with observations, as tight limits on spatial curvature \cite{Clarkson:2011gm, Okouma:2012dy, Ade:2015xua} only apply if it evolves like $\av{\sR}\propto(1+z)^2$, and in this case $\av{\sR}$ falls more rapidly to the past, and therefore has a smaller effect on the angular diameter distance \cite{Boehm:2013qqa}.} Combining the measurement $H_0=72.5\pm3.2$ km/s/Mpc \cite{Zhang:2017aqn} (for other determinations of $H_0$, see \cite{Efstathiou:2013via, *Riess:2016jrr, *Cardona:2016ems, *Follin:2017ljs, *Feeney:2017sgx}) with a model-independent determination of $t_0$ \cite{Krauss:2003} gives  $H_0 t_0=0.99\pm0.04$ for the central value $t_0=13.4$ Gyr, and $H_0t_0>0.83$ for the central value of $H_0$ and the lower limit $t_0>11.2$ Gyr.
Both of these agreements were noted in \cite{Rasanen:2008it} for the spherical collapse model.

The orders of magnitude are easy to understand. The Einstein--de Sitter universe has $Ht=2/3$ and a universe without matter and dominated by spatial curvature has $Ht=1$. Therefore, if the volume is dominated by voids but not completely empty, $2/3<Ht<1$.
The timing comes from the fact that the transfer function rises when approaching $\keq$ from above, so when modes with smaller wavenumber form non-linear structures, the number density of troughs grows, as shown in \fig{fig:volfrac}. After modes with $k\sim\keq$ collapse, the transfer function is practically constant, as shown in \fig{fig:Tk}. As the change in the transfer function is not sharp and peaks have only a small impact, the change in $Ht$ is not sharp.

Indeed, the matter-equality scale is the only scale that has a large impact on structure formation, apart from the free-streaming cut-off at large wavenumbers, which determines the beginning of structure formation \cite{Hofmann:2002nu, *Green:2003un, *Green:2005fa, *Green:2005kf} (the baryon acoustic oscillation scale, for example, is only a small correction). The time when modes with wavenumber $\keq$ form structures is related to the matter-radiation equality time $\teq=5\times10^4$ yr roughly as $t\sim A^{-3/2}\teq\sim10^{2}$ Gyr, where $A=5\times10^{-5}$ \cite{Ade:2015xua} is the primordial amplitude of perturbations \cite{Rasanen:2008it}. With the relation between $t$ and $R$ given by $\sigma_0(R,t)=1$, the age at late times is exponentially sensitive to the value of $R$, so we should avoid reading too much into the feature that the large values roughly corresponding to present-day observations are reached for times that are a few orders of magnitude larger than the than observed timing of the acceleration in the real universe. The fact that the relevant timescale for the start of significant backreaction is billions of years is robust. There is thus no coincidence problem, and anthropic arguments for the acceleration timescale related to the existence of highly evolved bound structures needed to produce observers like us are reduced to the change in the expansion rate coinciding with a late period in structure formation.

The details of the expansion history are different from the observations. This may not be too worrisome, given the simplicity of the model, which is missing many physical effects. Our treatment of structure formation is rather simplistic, as shown by the fact that half of the mass remains in underdense regions. The feature that smaller peaks may reside inside larger peaks and that troughs can be extinguished by peaks, known as the cloud-in-a-cloud and the void-in-a-cloud problem, respectively, is not included; for discussion in the context of the peaks formalism, see \cite{Paranjape:2012ks, Paranjape:2012jt}. Taking this into account would transfer mass from underdense to overdense regions. Another problem is that peaks and troughs are assumed to contain the same mass regardless of depth.
Also, the merging together of structures is dealt with only via the Gaussian smoothing, and the constraint $\sigma_0(R,t)=1$ is rather approximate. The condition that all matter is in troughs and peaks at late times is somewhat arbitrary, and only strictly works if the spectrum is scale-invariant; for a red spectrum, the peak number density keeps rising.
The treatment of the individual regions as either anisotropic but homogeneous (in the ellipsoidal model) or inhomogeneous but isotropic (in the spherical model) could also be improved. One possibility would be to use an exact general relativistic model, such as the Szekeres model \cite{Szekeres:1975}, the most general known dust solution. However, if holes are removed from a FRW universe and filled with a regular Szekeres dust model, backreaction is small if the holes are small compared to the Hubble scale and the matching between the hole and the background is taken into account \cite{Lavinto:2013exa}.
But the description of matter as dust breaks down due to shell crossings in the course of structure formation, and rotation also becomes non-negligible \cite{Reischke:2016eza}. Also, the Szekeres model is limited in the sense that the magnetic part of the Weyl tensor is zero, so there is no exchange of information between worldlines comoving with the dust; it is conjectured that generalisation of the model would violate this \cite{vanElst:1996zs, *Sopuerta:1997}. Vanishing magnetic part of the Weyl tensor is a property shared by Newtonian gravity, whereas it is important for real structure formation \cite{Kofman:1994pz, Ellis:1971pg, *Ellis:1994md, *Matarrese:1995sb, *vanElst:1998kb, *Ehlers:1997, *Ehlers:2009uv}. In the Newtonian limit, the magnetic part of the Weyl tensor is related to spatial curvature \cite{Rasanen:2011bm} (see also \cite{Clifton:2016mxx}).

Apart from issues with modelling the individual regions, the ensemble treatment has consistency problems. We have calculated the expansion rate $H$, the variance $\Delta\theta$ and the shear $\av{\sigma^2}$ from the ensemble. We have then determined $\OQ$ from its definition using these quantities. The quantities $q$ and $\Om$ have been calculated from the Buchert equations with $H$, using the first and second equality of \re{q}. We would get roughly the same answer if we were to use the definition $\Om\propto\av{\rho}/H^2$, taking $\av{\rho}$ to be inversely proportional to the volume calculated from the ensemble average used as the volume factor in \re{statH} and then using the second equality of \re{q}. The resulting $q$ is shown in \fig{fig:q} in black.\footnote{In \cite{Rasanen:2008it}, the analogue between a spherical Newtonian model and the relativistic dust FRW model was used to calculate $\OR$, and the matter density parameter was then determined as $\Om=1-\OR-\OQ$. This gives equivalent results in the spherical case, but there is no such simple analogue in the ellipsoidal case.}

The problem is that this volume does not agree with the volume
calculated from the expansion given by $H$, which is
$\propto e^{3\int\rmd t H(t)}$. As the expansion rate is faster than
in the Einstein--de Sitter case, the corresponding $\Om$ decreases
without limit. The deceleration parameter $q$ and the density
parameters determined by calculating $\Om$ (instead of $q$) from the
expansion rate $Ht$ are shown in figures \ref{fig:q} (black line) and
\ref{fig:omegas2}.  This is physically reasonable: when the volume
becomes dominated by very empty regions, the variance and shear become
small, so the evolution should tend to the FRW universe dominated by
negative spatial curvature, with $\Om$ approaching zero. Backreaction
can interpolate between two FRW cases, but it cannot continue to be
important when the universe becomes smooth. (Non-negligible residual
variance and shear could asymptotically persist due to ongoing
structure formation, so that the evolution does not asymptotically
approach $Ht=1$, but $\Om$ would nevertheless approach zero if the
expansion is faster than in the Einstein--de Sitter case.)

The form of the Buchert equations \re{Ray}--\re{cons} and the resulting density parameter relations \re{q} and \re{omegas} that we have used rely on the matter evolving like irrotational dust. Virialisation and our treatment of the merging of structures violate this assumption, so there are terms missing from the equations. However, the above inconsistency cannot be laid at the door of the dust assumption.
The root of the problem is that we consider an ensemble of regions with different expansion rates, not an ensemble of regions with different volumes. More precisely, the expansion rate we consider involves the time derivative of the factor $s$ in \re{statH} that accounts for the growth factor for individual regions, but not $f$, which models the merging together of regions.
Simply put, if $Ht$ is constant and $>2/3$, then $\Om$ cannot be a non-zero constant, so calculating them both directly using the ensemble quantities that practically saturate after $\keq R\approx9$ (or equivalently $t\approx 4\times10^{5}$ Gyr) is not consistent.

A related issue is that the structures evolve in the background space, unaffected by the deviation of the expansion rate from the background. A small aspect of this is that when the expansion becomes faster, the growth of density perturbations should change. A potentially more important factor is that the volume factor in the ensemble average \re{statH} for regions that have stopped evolving, \ie ones corresponding to $k^{-1}\ll R$, is equal to the background volume. However, even if the structures on small scales have stopped evolving, their volume should reflect the extra expansion that has taken place relative to the background. Correcting the volume factor should also take into account the discrepancy between the ensemble volume and the volume determined from the expansion rate.
However, the usefulness of such improvements is conditional on making sure that the distinction between Newtonian and relativistic behaviour is correctly implemented in the statistical treatment.

\para{Newtonian and relativistic treatment.}

Our calculation looks completely Newtonian. However, in the Newtonian case the contribution of the variance and the shear to $\sQ$ cancel each other up to a boundary term \cite{Buchert:1995fz}. For a statistically homogeneous and isotropic distribution, the boundary term is small, so the expansion rate is close to the FRW value, whereas we have found significant differences. It was noted in \cite{Rasanen:2008it} that the resolution is that in the Newtonian case total energy is conserved, and we have not implemented such a global constraint on the peak statistics.
In contrast, in the relativistic case the total energy is not a well-defined quantity. The corresponding term in the equations for the average expansion rate is the spatial curvature, for which there is no conservation law: there is no requirement for the positive spatial curvature of overdense regions to exactly balance the negative spatial curvature of underdense regions in the course of evolution \cite{Buchert:2011sx, Buchert:2017obp}.
It was therefore argued in \cite{Rasanen:2008it} that a calculation like the one presented here is closer to the relativistic situation: even if the individual regions are Newtonian, their distribution does not follow the Newtonian pattern. Of course, in a full and consistent calculation, the distribution of regions should flow from the process of regions joining together, which in our calculation is modelled only by changing the smoothing scale.

Let us consider the energy argument more carefully. In Newtonian theory, the total energy of a distribution of self-gravitating dust is (see \eg \cite{Bond:1996})\footnote{See also \cite{Gibbons:2013msa} for a study of discrete Newtonian cosmology.}
\bea \label{E}
  E &=& \frac{1}{2} \sum_i m_i \dot{\bar{x_i}}^2 - \GN \sum_{i<j} \frac{m_i m_j}{|\bar{x}_i-\bar{x}_j|} \el
  &=& \frac{1}{2} \int\rmd^3 x \rho(t,\bar x) \bar{v}(t,\bar{x})^2 - \frac{1}{2} \GN \int\rmd^3 x \int\rmd^3 y \frac{\rho(t,\bar x) \rho(t,\bar y)}{|\bar{x}-\bar{y}|} \ ,
\eea
where on the first line we have point particles with masses $m_i$ and positions $\bar x_i$, and on the second line we have gone over to a description in terms of a continuous fluid with density $\rho(t,\bar x)$ and velocity $\bar v$. For a homogeneous and isotropic region with total mass $M=\frac{4\pi}{3}a(t)^3\rho(t)$, where $a(t)$ is the radius, this reduces to \cite{Peebles:1980} (page 86)
\bea \label{E2}
  E = \frac{M}{10} a^2 \left( 3 \frac{\adot^2}{a^2} - 8\pi\GN\rho \right) \ ,
\eea
which is conserved. If we consider a dust FRW model with spatial curvature $\sR=-6 K a^{-2}$, then \re{E2} corresponds to the first Friedmann equation with the identification $E=-\frac{3}{10} M K$. So for this simple system, conservation of the Newtonian energy does correspond to the FRW spatial curvature evolving like $\propto a^{-2}$, in agreement with the above argument.

What about matter distribution that is not homogeneous and isotropic? With $\sQ=0$, the average Raychaudhuri equation \re{Ray} has (applying \re{cons}) the same form as in the homogeneous and isotropic case.\footnote{The Raychaudhuri equation \re{Ray} is written for the relativistic case, but the Newtonian form is exactly the same.} Therefore its first integral gives again the first Friedmann equation with a constant of integration like \re{E2}.\footnote{The equation \re{Ray}  has the form of a Newtonian one-dimensional force law, so for any $\sQ=\sQ(a)$ it yields a conserved quantity. However, in general a non-zero $\sQ$ is not a function of the scale factor alone.}
In contrast, in the relativistic case, equation \re{Ham} that generalises the first Friedmann equation is an independent equation, and its integrability condition with \re{Ray} shows that in general $a^2 \av{\sR}$ is not conserved, as $\sQ\neq0$. This corresponds to the fact that the conservation law for the spatial curvature in the relativistic FRW model is due to homogeneity and isotropy; in contrast the Newtonian case energy is conserved regardless of the symmetry (or lack of it) of the matter distribution.
However, this energy explanation is limited in that conservation of the Newtonian energy does not, in fact, generally imply that backreaction vanishes, $\sQ=0$. For a single homogeneous non-spherical ellipsoid embedded in empty space, the Newtonian energy is conserved, but $\sQ\neq0$ (the variance is zero and the shear is non-zero). Another example is the union of two disjoint spherical regions. The energy of each region is conserved, so the total energy is also conserved, but $\sQ\neq0$, as energy is additive, but $\sQ$ is not \cite{Rasanen:2006zw, Rasanen:2006kp}. It is not clear whether the Newtonian energy can always be associated with the conserved quantity coming from the Raychaudhuri equation in the cases when $\sQ=0$.

In some studies backreaction has been estimated by combining statistical modelling and Newtonian $N$-body simulation data in various degrees \cite{Wiegand:2010uh, *Wiegand:2011gs, *Roukema:2013cya, *Racz:2016rss, *Roukema:2017doi}. However, because backreaction on the expansion rate is zero for Newtonian gravity with periodic boundary conditions \cite{Buchert:1995fz}, non-zero backreaction is due to neglect of spherical asymmetry and/or proper matching of the regions to each other and the environment (see \cite{Kaiser:2017hqn, Buchert:2017obp} for recent discussion). While such studies may be useful in demonstrating timescales and orders of magnitude, there is limited information to be gained from any analysis that does not properly account for the difference between Newtonian and relativistic degrees of freedom and constraints in the non-linear regime.

\section{Conclusions} \label{sec:conc}

\para{The effect of structures.}

We have considered a statistical model for the effect of structure formation on the average expansion rate. We sample the initial Gaussian density field for ellipsoidal peaks and troughs with a time-dependent smoothing scale $R(t)$ determined by setting the root mean square density contrast to unity at all times, $\sigma_0(R,t)=1$, so that we look at the generation of typical structures that is forming at each era. The peaks and troughs are modelled as homogeneous Newtonian ellipsoids, with the volume that is not in peaks and troughs taken to expand like the background. This extends the calculation of \cite{Rasanen:2008it}, where spherical regions were considered. We have also made other small improvements, like using spectral index $n=0.96$ rather than the scale-invariant case, and using the numerical transfer function from CAMB. The ellipsoidal model allows us to estimate the contribution of the shear and the fraction of regions that have filamentary or planar structure. Our modelling does not properly capture the transfer of mass from underdense to overdense regions, so only half of the total mass is in overdense regions, in contrast to simulations and observations \cite{Bond:1995yt, Cautun:2014fwa}.
Therefore filaments, sheets and clusters do not have a large effect in our calculation, the most important feature is the growth of the underdense regions.

Like the spherical case, the ellipsoidal case shows an increase of the expansion rate of the right order of magnitude,  compared to observations, at late times. In the ellipsoidal case, $Ht$ rises from the Einstein--de Sitter value $Ht=2/3$, saturating at $0.83$ around $t=10^5$ Gyr. The shear slows down the expansion rate, so this is smaller than the spherical case value $Ht=0.85$. This number is sensitive to the choice of smoothing scale. Changing the smoothing scale from $\sigma_0(R,t)=1$ to 0.5 or 2 gives 0.74 or 0.88 as the asymptotic value.
There is no acceleration, the expansion just decelerates more slowly. The change in the expansion rate is due to underdense regions expanding more rapidly than the background, so $Ht$ is naturally between 2/3 and the completely empty case value $Ht=1$. The timing comes from the fact that the CDM transfer function rises around $\keq$, so the number density of peaks and troughs grows when modes with $k\sim\keq$ enter the non-linear regime. The timescale for  this is determined by the matter-radiation equality time $\teq=5\times10^4$ yr and the small primordial perturbation amplitude $A=5\times10^{-5}$ as $t\sim A^{-3/2}\teq\sim10^{2}$ Gyr.

\para{Open questions.}

It is non-trivial that the right order of magnitude in the amplitude and roughly right timescale of the change in the expansion rate follow simply from the known physics of structure formation. However, the model has shortcomings that would need to be overcome for the results to be more than suggestive. All dimensionless quantities calculated from the ensemble are almost constant after structures with wavenumber $\keq$ collapse, because the transfer function on those scales and the primordial spectrum are close to scale-invariant. However, as the expansion is faster than in the Einstein--de Sitter case, $\Om$ should approach zero, not remain constant.
This issue could be resolved with a more careful statistical treatment. A more difficult problem is whether the sampling of regions correctly reproduces the distinction between relativistic and Newtonian physics. This is a crucial question, given that backreaction is always small in a statistically homogeneous and isotropic universe in the Newtonian case, but not in general relativity \cite{Buchert:1995fz, Buchert:1999mc, Lavinto:2013exa}. Studies of backreaction that use data from Newtonian $N$-body simulations \cite{Wiegand:2010uh, *Wiegand:2011gs, *Roukema:2013cya, *Racz:2016rss, *Roukema:2017doi} face the same issue of how to relate the local Newtonian approximation to the global general relativistic setting.
It remains to be seen whether this question can be answered via improved statistical models or if it will only be settled by relativistic cosmological simulations \cite{Adamek:2013wja, *Adamek:2014gva, *Adamek:2014xba, *Adamek:2015eda, *Adamek:2016zes, Yoo:2012jz, *Bentivegna:2012ei, *Bentivegna:2013ata, *Adamek:2015hqa, *Bentivegna:2016fls, bentivegna:2015flc, *Giblin:2015vwq, *Mertens:2015ttp, *Macpherson:2016ict, *Giblin:2017juu}.

\acknowledgments

We thank Shaun Hotchkiss for contributing to this work.

\bibliographystyle{JHEPM}
\bibliography{refs}

\ifx\mcitethebibliography\mciteundefinedmacro
\PackageError{unsrtM.bst}{mciteplus.sty has not been loaded}
{This bibstyle requires the use of the mciteplus package.}\fi
\providecommand{\href}[2]{#2}\begingroup\raggedright\begin{mcitethebibliography}{100}

\bibitem{Bonnor1986}
W.~B. {Bonnor} and G.~F.~R. {Ellis}, \emph{{Observational homogeneity of the
  universe}}, \href{http://dx.doi.org/10.1093/mnras/218.4.605}{\emph{\mnras}
  {\bf 218} (Feb., 1986) 605--614}.

\bibitem{Stoeger1987}
W.~R. {Stoeger}, G.~F.~R. {Ellis} and C.~{Hellaby}, \emph{{The relationship
  between continuum homogeneity and statistical homogeneity in cosmology}},
  \href{http://dx.doi.org/10.1093/mnras/226.2.373}{\emph{\mnras} {\bf 226}
  (May, 1987) 373--381}.

\bibitem{Clarkson:2010uz}
C.~Clarkson and R.~Maartens, \emph{{Inhomogeneity and the foundations of
  concordance cosmology}},
  \href{http://dx.doi.org/10.1088/0264-9381/27/12/124008}{\emph{Class. Quant.
  Grav.} {\bf 27} (2010) 124008}, [\href{http://arxiv.org/abs/1005.2165}{{\tt
  1005.2165}}].

\bibitem{Heavens:2011mr}
A.~F. Heavens, R.~Jimenez and R.~Maartens, \emph{{Testing homogeneity with the
  fossil record of galaxies}},
  \href{http://dx.doi.org/10.1088/1475-7516/2011/09/035}{\emph{JCAP} {\bf 1109}
  (2011) 035}, [\href{http://arxiv.org/abs/1107.5910}{{\tt 1107.5910}}].

\bibitem{Clifton:2011sn}
T.~Clifton, C.~Clarkson and P.~Bull, \emph{{The isotropic blackbody CMB as
  evidence for a homogeneous universe}},
  \href{http://dx.doi.org/10.1103/PhysRevLett.109.051303}{\emph{Phys. Rev.
  Lett.} {\bf 109} (2012) 051303}, [\href{http://arxiv.org/abs/1111.3794}{{\tt
  1111.3794}}].

\bibitem{Hoyle:2012pb}
B.~Hoyle, R.~Tojeiro, R.~Jimenez, A.~Heavens, C.~Clarkson and R.~Maartens,
  \emph{{Testing homogeneity with galaxy star formation history}},
  \href{http://dx.doi.org/10.1088/2041-8205/762/1/L9}{\emph{Astrophys. J.} {\bf
  762} (2012) L9}, [\href{http://arxiv.org/abs/1209.6181}{{\tt 1209.6181}}].

\bibitem{Hogg:2004vw}
D.~W. Hogg, D.~J. Eisenstein, M.~R. Blanton, N.~A. Bahcall, J.~Brinkmann, J.~E.
  Gunn et~al., \emph{{Cosmic homogeneity demonstrated with luminous red
  galaxies}}, \href{http://dx.doi.org/10.1086/429084}{\emph{Astrophys. J.} {\bf
  624} (2005) 54--58}, [\href{http://arxiv.org/abs/astro-ph/0411197}{{\tt
  astro-ph/0411197}}].

\bibitem{Labini:2009zi}
F.~S. Labini, \emph{{Characterizing the large scale inhomogeneity of the galaxy
  distribution}}, \href{http://dx.doi.org/10.1063/1.3462744}{\emph{AIP Conf.
  Proc.} {\bf 1241} (2010) 981--990},
  [\href{http://arxiv.org/abs/0910.3833}{{\tt 0910.3833}}].

\bibitem{Labini:2010aj}
F.~S. Labini and L.~Pietronero, \emph{{The complex universe: recent
  observations and theoretical challenges}},
  \href{http://dx.doi.org/10.1088/1742-5468/2010/11/P11029}{\emph{J. Stat.
  Mech.} {\bf 1011} (2010) P11029}, [\href{http://arxiv.org/abs/1012.5624}{{\tt
  1012.5624}}].

\bibitem{Labini:2011tj}
F.~S. Labini, \emph{{Inhomogeneities in the universe}},
  \href{http://dx.doi.org/10.1088/0264-9381/28/16/164003}{\emph{Class. Quant.
  Grav.} {\bf 28} (2011) 164003}, [\href{http://arxiv.org/abs/1103.5974}{{\tt
  1103.5974}}].

\bibitem{Labini:2011dv}
F.~S. Labini, \emph{{Very large scale correlations in the galaxy
  distribution}},
  \href{http://dx.doi.org/10.1209/0295-5075/96/59001}{\emph{EPL} {\bf 96}
  (2011) 59001}, [\href{http://arxiv.org/abs/1110.4041}{{\tt 1110.4041}}].

\bibitem{Scrimgeour:2012wt}
M.~Scrimgeour et~al., \emph{{The WiggleZ Dark Energy Survey: the transition to
  large-scale cosmic homogeneity}},
  \href{http://dx.doi.org/10.1111/j.1365-2966.2012.21402.x}{\emph{Mon. Not.
  Roy. Astron. Soc.} {\bf 425} (2012) 116--134},
  [\href{http://arxiv.org/abs/1205.6812}{{\tt 1205.6812}}].

\bibitem{Nadathur:2013mva}
S.~Nadathur, \emph{{Seeing patterns in noise: Gigaparsec-scale `structures'
  that do not violate homogeneity}},
  \href{http://dx.doi.org/10.1093/mnras/stt1028}{\emph{Mon. Not. Roy. Astron.
  Soc.} {\bf 434} (2013) 398--406}, [\href{http://arxiv.org/abs/1306.1700}{{\tt
  1306.1700}}].

\bibitem{Labini:2014zoa}
F.~S. Labini, D.~Tekhanovich and Y.~V. Baryshev, \emph{{Spatial density
  fluctuations and selection effects in galaxy redshift surveys}},
  \href{http://dx.doi.org/10.1088/1475-7516/2014/07/035}{\emph{JCAP} {\bf 1407}
  (2014) 035}, [\href{http://arxiv.org/abs/1406.5899}{{\tt 1406.5899}}].

\bibitem{Laurent:2016eqo}
P.~Laurent et~al., \emph{{A 14 $h^{-3}$ Gpc$^3$ study of cosmic homogeneity
  using BOSS DR12 quasar sample}},
  \href{http://dx.doi.org/10.1088/1475-7516/2016/11/060}{\emph{JCAP} {\bf 1611}
  (2016) 060}, [\href{http://arxiv.org/abs/1602.09010}{{\tt 1602.09010}}].

\bibitem{Ntelis:2017nrj}
P.~Ntelis et~al., \emph{{Exploring cosmic homogeneity with the BOSS DR12 galaxy
  sample}}, \href{http://dx.doi.org/10.1088/1475-7516/2017/06/019}{\emph{JCAP}
  {\bf 1706} (2017) 019}, [\href{http://arxiv.org/abs/1702.02159}{{\tt
  1702.02159}}].

\bibitem{Rasanen:2009mg}
S.~Rasanen, \emph{{On the relation between the isotropy of the CMB and the
  geometry of the universe}},
  \href{http://dx.doi.org/10.1103/PhysRevD.79.123522}{\emph{Phys. Rev.} {\bf
  D79} (2009) 123522}, [\href{http://arxiv.org/abs/0903.3013}{{\tt
  0903.3013}}].

\bibitem{Maartens:2011yx}
R.~Maartens, \emph{{Is the Universe homogeneous?}},
  \href{http://dx.doi.org/10.1098/rsta.2011.0289}{\emph{Phil. Trans. Roy. Soc.
  Lond.} {\bf A369} (2011) 5115--5137},
  [\href{http://arxiv.org/abs/1104.1300}{{\tt 1104.1300}}].

\bibitem{Ellis:1984bqf}
G.~F.~R. Ellis, \emph{{Relativistic Cosmology: Its Nature, Aims and Problems}},
  \href{http://dx.doi.org/10.1007/978-94-009-6469-3_14}{\emph{Fundam. Theor.
  Phys.} {\bf 9} (1984) 215--288}.

\bibitem{Ellis:1987zz}
G.~F.~R. Ellis and W.~Stoeger, \emph{{The 'fitting problem' in cosmology}},
  \href{http://dx.doi.org/10.1088/0264-9381/4/6/025}{\emph{Class. Quant. Grav.}
  {\bf 4} (1987) 1697--1729}.

\bibitem{Ellis:2005uz}
G.~F.~R. Ellis and T.~Buchert, \emph{{The Universe seen at different scales}},
  \href{http://dx.doi.org/10.1016/j.physleta.2005.06.087}{\emph{Phys. Lett.}
  {\bf A347} (2005) 38--46}, [\href{http://arxiv.org/abs/gr-qc/0506106}{{\tt
  gr-qc/0506106}}].

\bibitem{Rasanen:2011ki}
S.~Rasanen, \emph{{Backreaction: directions of progress}},
  \href{http://dx.doi.org/10.1088/0264-9381/28/16/164008}{\emph{Class. Quant.
  Grav.} {\bf 28} (2011) 164008}, [\href{http://arxiv.org/abs/1102.0408}{{\tt
  1102.0408}}].

\bibitem{Buchert:2011sx}
T.~Buchert and S.~Räsänen, \emph{{Backreaction in late-time cosmology}},
  \href{http://dx.doi.org/10.1146/annurev.nucl.012809.104435}{\emph{Ann. Rev.
  Nucl. Part. Sci.} {\bf 62} (2012) 57--79},
  [\href{http://arxiv.org/abs/1112.5335}{{\tt 1112.5335}}].

\bibitem{Buchert:1999mc}
T.~Buchert, \emph{{On average properties of inhomogeneous cosmologies}},  in
  \emph{{Proceedings, 9th Workshop on General Relativity and Gravitation
  (JGRG9): Hiroshima, Japan, November 3-6, 1999}}, 1999.
\newblock \href{http://arxiv.org/abs/gr-qc/0001056}{{\tt gr-qc/0001056}}.

\bibitem{Wetterich:2001kr}
C.~Wetterich, \emph{{Can structure formation influence the cosmological
  evolution?}}, \href{http://dx.doi.org/10.1103/PhysRevD.67.043513}{\emph{Phys.
  Rev.} {\bf D67} (2003) 043513},
  [\href{http://arxiv.org/abs/astro-ph/0111166}{{\tt astro-ph/0111166}}].

\bibitem{Schwarz:2002ba}
D.~J. Schwarz, \emph{{Accelerated expansion without dark energy}},  in
  \emph{{18th IAP Colloquium on the Nature of Dark Energy: Observational and
  Theoretical Results on the Accelerating Universe Paris, France, July 1-5,
  2002}}, 2002.
\newblock \href{http://arxiv.org/abs/astro-ph/0209584}{{\tt astro-ph/0209584}}.

\bibitem{Rasanen:2003fy}
S.~Rasanen, \emph{{Dark energy from backreaction}},
  \href{http://dx.doi.org/10.1088/1475-7516/2004/02/003}{\emph{JCAP} {\bf 0402}
  (2004) 003}, [\href{http://arxiv.org/abs/astro-ph/0311257}{{\tt
  astro-ph/0311257}}].

\bibitem{Rasanen:2004sa}
S.~Rasanen, \emph{{Backreaction of linear perturbations and dark energy}},  in
  \emph{{39th Rencontres de Moriond Workshop on Exploring the Universe:
  Contents and Structures of the Universe La Thuile, Italy, March 28-April 4,
  2004}}, 2004.
\newblock \href{http://arxiv.org/abs/astro-ph/0407317}{{\tt astro-ph/0407317}}.

\bibitem{Buchert:1995fz}
T.~Buchert and J.~Ehlers, \emph{{Averaging inhomogeneous Newtonian
  cosmologies}}, {\emph{Astron. Astrophys.} {\bf 320} (1997) 1--7},
  [\href{http://arxiv.org/abs/astro-ph/9510056}{{\tt astro-ph/9510056}}].

\bibitem{Enqvist:2009hn}
K.~Enqvist, M.~Mattsson and G.~Rigopoulos, \emph{{Supernovae data and
  perturbative deviation from homogeneity}},
  \href{http://dx.doi.org/10.1088/1475-7516/2009/09/022}{\emph{JCAP} {\bf 0909}
  (2009) 022}, [\href{http://arxiv.org/abs/0907.4003}{{\tt 0907.4003}}].

\bibitem{Green:2010qy}
S.~R. Green and R.~M. Wald, \emph{{A new framework for analyzing the effects of
  small scale inhomogeneities in cosmology}},
  \href{http://dx.doi.org/10.1103/PhysRevD.83.084020}{\emph{Phys. Rev.} {\bf
  D83} (2011) 084020}, [\href{http://arxiv.org/abs/1011.4920}{{\tt
  1011.4920}}].

\bibitem{Green:2013yua}
S.~R. Green and R.~M. Wald, \emph{{Examples of backreaction of small scale
  inhomogeneities in cosmology}},
  \href{http://dx.doi.org/10.1103/PhysRevD.87.124037}{\emph{Phys. Rev.} {\bf
  D87} (2013) 124037}, [\href{http://arxiv.org/abs/1304.2318}{{\tt
  1304.2318}}].

\bibitem{Green:2014aga}
S.~R. Green and R.~M. Wald, \emph{{How well is our universe described by an
  FLRW model?}},
  \href{http://dx.doi.org/10.1088/0264-9381/31/23/234003}{\emph{Class. Quant.
  Grav.} {\bf 31} (2014) 234003}, [\href{http://arxiv.org/abs/1407.8084}{{\tt
  1407.8084}}].

\bibitem{Buchert:2015iva}
T.~Buchert et~al., \emph{{Is there proof that backreaction of inhomogeneities
  is irrelevant in cosmology?}},
  \href{http://dx.doi.org/10.1088/0264-9381/32/21/215021}{\emph{Class. Quant.
  Grav.} {\bf 32} (2015) 215021}, [\href{http://arxiv.org/abs/1505.07800}{{\tt
  1505.07800}}].

\bibitem{Green:2015bma}
S.~R. Green and R.~M. Wald, \emph{{Comments on Backreaction}},
  \href{http://arxiv.org/abs/1506.06452}{{\tt 1506.06452}}.

\bibitem{Ostrowski:2015pzb}
J.~J. Ostrowski and B.~F. Roukema, \emph{{On the Green and Wald formalism}},
  in \emph{{14th Marcel Grossmann Meeting on Recent Developments in Theoretical
  and Experimental General Relativity, Astrophysics, and Relativistic Field
  Theories (MG14) Rome, Italy, July 12-18, 2015}}, 2015.
\newblock \href{http://arxiv.org/abs/1512.02947}{{\tt 1512.02947}}.

\bibitem{Green:2016cwo}
S.~R. Green and R.~M. Wald, \emph{{A simple, heuristic derivation of our ‘no
  backreaction’ results}},
  \href{http://dx.doi.org/10.1088/0264-9381/33/12/125027}{\emph{Class. Quant.
  Grav.} {\bf 33} (2016) 125027}, [\href{http://arxiv.org/abs/1601.06789}{{\tt
  1601.06789}}].

\bibitem{Lavinto:2013exa}
M.~Lavinto, S.~Räsänen and S.~J. Szybka, \emph{{Average expansion rate and
  light propagation in a cosmological Tardis spacetime}},
  \href{http://dx.doi.org/10.1088/1475-7516/2013/12/051}{\emph{JCAP} {\bf 1312}
  (2013) 051}, [\href{http://arxiv.org/abs/1308.6731}{{\tt 1308.6731}}].

\bibitem{Clarkson:2007pz}
C.~Clarkson, B.~Bassett and T.~H.-C. Lu, \emph{{A general test of the
  Copernican Principle}},
  \href{http://dx.doi.org/10.1103/PhysRevLett.101.011301}{\emph{Phys. Rev.
  Lett.} {\bf 101} (2008) 011301}, [\href{http://arxiv.org/abs/0712.3457}{{\tt
  0712.3457}}].

\bibitem{Rasanen:2013swa}
S.~Räsänen, \emph{{A covariant treatment of cosmic parallax}},
  \href{http://dx.doi.org/10.1088/1475-7516/2014/03/035}{\emph{JCAP} {\bf 1403}
  (2014) 035}, [\href{http://arxiv.org/abs/1312.5738}{{\tt 1312.5738}}].

\bibitem{Rasanen:2014mca}
S.~Räsänen, K.~Bolejko and A.~Finoguenov, \emph{{New Test of the
  Friedmann-Lemaître-Robertson-Walker Metric Using the Distance Sum Rule}},
  \href{http://dx.doi.org/10.1103/PhysRevLett.115.101301}{\emph{Phys. Rev.
  Lett.} {\bf 115} (2015) 101301}, [\href{http://arxiv.org/abs/1412.4976}{{\tt
  1412.4976}}].

\bibitem{Shafieloo:2009hi}
A.~Shafieloo and C.~Clarkson, \emph{{Model independent tests of the standard
  cosmological model}},
  \href{http://dx.doi.org/10.1103/PhysRevD.81.083537}{\emph{Phys. Rev.} {\bf
  D81} (2010) 083537}, [\href{http://arxiv.org/abs/0911.4858}{{\tt
  0911.4858}}].

\bibitem{Mortsell:2011yk}
E.~Mortsell and J.~Jonsson, \emph{{A model independent measure of the large
  scale curvature of the Universe}},
  \href{http://arxiv.org/abs/1102.4485}{{\tt 1102.4485}}.

\bibitem{Sapone:2014nna}
D.~Sapone, E.~Majerotto and S.~Nesseris, \emph{{Curvature versus distances:
  Testing the FLRW cosmology}},
  \href{http://dx.doi.org/10.1103/PhysRevD.90.023012}{\emph{Phys. Rev.} {\bf
  D90} (2014) 023012}, [\href{http://arxiv.org/abs/1402.2236}{{\tt
  1402.2236}}].

\bibitem{Cai:2015pia}
R.-G. Cai, Z.-K. Guo and T.~Yang, \emph{{Null test of the cosmic curvature
  using $H(z)$ and supernovae data}},
  \href{http://dx.doi.org/10.1103/PhysRevD.93.043517}{\emph{Phys. Rev.} {\bf
  D93} (2016) 043517}, [\href{http://arxiv.org/abs/1509.06283}{{\tt
  1509.06283}}].

\bibitem{LHuillier:2016mtc}
B.~L'Huillier and A.~Shafieloo, \emph{{Model-independent test of the FLRW
  metric, the flatness of the Universe, and non-local measurement of
  $H_0r_\mathrm{d}$}},
  \href{http://dx.doi.org/10.1088/1475-7516/2017/01/015}{\emph{JCAP} {\bf 1701}
  (2017) 015}, [\href{http://arxiv.org/abs/1606.06832}{{\tt 1606.06832}}].

\bibitem{Yu:2016gmd}
H.~Yu and F.~Y. Wang, \emph{{New model-independent method to test the curvature
  of the universe}},
  \href{http://dx.doi.org/10.3847/0004-637X/828/2/85}{\emph{Astrophys. J.} {\bf
  828} (2016) 85}, [\href{http://arxiv.org/abs/1605.02483}{{\tt 1605.02483}}].

\bibitem{Li:2016wjm}
Z.~Li, G.-J. Wang, K.~Liao and Z.-H. Zhu, \emph{{Model-independent estimations
  for the curvature from standard candles and clocks}},
  \href{http://dx.doi.org/10.3847/1538-4357/833/2/240}{\emph{Astrophys. J.}
  {\bf 833} (2016) 240}, [\href{http://arxiv.org/abs/1611.00359}{{\tt
  1611.00359}}].

\bibitem{Wei:2016xti}
J.-J. Wei and X.-F. Wu, \emph{{An Improved Method to Measure the Cosmic
  Curvature}},
  \href{http://dx.doi.org/10.3847/1538-4357/aa674b}{\emph{Astrophys. J.} {\bf
  838} (2017) 160}, [\href{http://arxiv.org/abs/1611.00904}{{\tt 1611.00904}}].

\bibitem{Montanari:2017yma}
F.~Montanari and S.~Rasanen, \emph{{Backreaction and FRW consistency
  conditions}},  \href{http://arxiv.org/abs/1709.06022}{{\tt 1709.06022}}.

\bibitem{Boehm:2013qqa}
C.~Boehm and S.~Räsänen, \emph{{Violation of the FRW consistency condition as
  a signature of backreaction}},
  \href{http://dx.doi.org/10.1088/1475-7516/2013/09/003}{\emph{JCAP} {\bf 1309}
  (2013) 003}, [\href{http://arxiv.org/abs/1305.7139}{{\tt 1305.7139}}].

\bibitem{Rasanen:2008be}
S.~Rasanen, \emph{{Light propagation in statistically homogeneous and isotropic
  dust universes}},
  \href{http://dx.doi.org/10.1088/1475-7516/2009/02/011}{\emph{JCAP} {\bf 0902}
  (2009) 011}, [\href{http://arxiv.org/abs/0812.2872}{{\tt 0812.2872}}].

\bibitem{Rasanen:2009uw}
S.~Rasanen, \emph{{Light propagation in statistically homogeneous and isotropic
  universes with general matter content}},
  \href{http://dx.doi.org/10.1088/1475-7516/2010/03/018}{\emph{JCAP} {\bf 1003}
  (2010) 018}, [\href{http://arxiv.org/abs/0912.3370}{{\tt 0912.3370}}].

\bibitem{Clarkson:2011br}
C.~Clarkson, G.~F.~R. Ellis, A.~Faltenbacher, R.~Maartens, O.~Umeh and J.-P.
  Uzan, \emph{{(Mis-)Interpreting supernovae observations in a lumpy
  universe}},
  \href{http://dx.doi.org/10.1111/j.1365-2966.2012.21750.x}{\emph{Mon. Not.
  Roy. Astron. Soc.} {\bf 426} (2012) 1121--1136},
  [\href{http://arxiv.org/abs/1109.2484}{{\tt 1109.2484}}].

\bibitem{Bull:2012zx}
P.~Bull and T.~Clifton, \emph{{Local and non-local measures of acceleration in
  cosmology}}, \href{http://dx.doi.org/10.1103/PhysRevD.85.103512}{\emph{Phys.
  Rev.} {\bf D85} (2012) 103512}, [\href{http://arxiv.org/abs/1203.4479}{{\tt
  1203.4479}}].

\bibitem{Clifton:2009jw}
T.~Clifton and P.~G. Ferreira, \emph{{Archipelagian Cosmology: Dynamics and
  Observables in a Universe with Discretized Matter Content}},
  \href{http://dx.doi.org/10.1103/PhysRevD.84.109902,
  10.1103/PhysRevD.80.103503}{\emph{Phys. Rev.} {\bf D80} (2009) 103503},
  [\href{http://arxiv.org/abs/0907.4109}{{\tt 0907.4109}}].

\bibitem{Clifton:2009bp}
T.~Clifton and P.~G. Ferreira, \emph{{Errors in Estimating $\Omega_\Lambda$ due
  to the Fluid Approximation}},
  \href{http://dx.doi.org/10.1088/1475-7516/2009/10/026}{\emph{JCAP} {\bf 0910}
  (2009) 026}, [\href{http://arxiv.org/abs/0908.4488}{{\tt 0908.4488}}].

\bibitem{Clifton:2012qh}
T.~Clifton, K.~Rosquist and R.~Tavakol, \emph{{An Exact quantification of
  backreaction in relativistic cosmology}},
  \href{http://dx.doi.org/10.1103/PhysRevD.86.043506}{\emph{Phys. Rev.} {\bf
  D86} (2012) 043506}, [\href{http://arxiv.org/abs/1203.6478}{{\tt
  1203.6478}}].

\bibitem{Clifton:2010fr}
T.~Clifton, \emph{{Cosmology Without Averaging}},
  \href{http://dx.doi.org/10.1088/0264-9381/28/16/164011}{\emph{Class. Quant.
  Grav.} {\bf 28} (2011) 164011}, [\href{http://arxiv.org/abs/1005.0788}{{\tt
  1005.0788}}].

\bibitem{Clifton:2011mt}
T.~Clifton, P.~G. Ferreira and K.~O'Donnell, \emph{{An Improved Treatment of
  Optics in the Lindquist-Wheeler Models}},
  \href{http://dx.doi.org/10.1103/PhysRevD.85.023502}{\emph{Phys. Rev.} {\bf
  D85} (2012) 023502}, [\href{http://arxiv.org/abs/1110.3191}{{\tt
  1110.3191}}].

\bibitem{Clifton:2013jpa}
T.~Clifton, D.~Gregoris, K.~Rosquist and R.~Tavakol, \emph{{Exact Evolution of
  Discrete Relativistic Cosmological Models}},
  \href{http://dx.doi.org/10.1088/1475-7516/2013/11/010}{\emph{JCAP} {\bf 1311}
  (2013) 010}, [\href{http://arxiv.org/abs/1309.2876}{{\tt 1309.2876}}].

\bibitem{Clifton:2014lha}
T.~Clifton, D.~Gregoris and K.~Rosquist, \emph{{Piecewise Silence in Discrete
  Cosmological Models}},
  \href{http://dx.doi.org/10.1088/0264-9381/31/10/105012}{\emph{Class. Quant.
  Grav.} {\bf 31} (2014) 105012}, [\href{http://arxiv.org/abs/1402.3201}{{\tt
  1402.3201}}].

\bibitem{Clifton:2014mza}
T.~Clifton, \emph{{The Method of Images in Cosmology}},
  \href{http://dx.doi.org/10.1088/0264-9381/31/17/175010}{\emph{Class. Quant.
  Grav.} {\bf 31} (2014) 175010}, [\href{http://arxiv.org/abs/1405.3197}{{\tt
  1405.3197}}].

\bibitem{Sanghai:2015wia}
V.~A.~A. Sanghai and T.~Clifton, \emph{{Post-Newtonian Cosmological
  Modelling}}, \href{http://dx.doi.org/10.1103/PhysRevD.93.089903,
  10.1103/PhysRevD.91.103532}{\emph{Phys. Rev.} {\bf D91} (2015) 103532},
  [\href{http://arxiv.org/abs/1503.08747}{{\tt 1503.08747}}].

\bibitem{Clifton:2015tra}
T.~Clifton, \emph{{What's the Matter in Cosmology?}},  2015.
\newblock \href{http://arxiv.org/abs/1509.06682}{{\tt 1509.06682}}.

\bibitem{Sanghai:2016ucv}
V.~A.~A. Sanghai and T.~Clifton, \emph{{Cosmological backreaction in the
  presence of radiation and a cosmological constant}},
  \href{http://dx.doi.org/10.1103/PhysRevD.94.023505}{\emph{Phys. Rev.} {\bf
  D94} (2016) 023505}, [\href{http://arxiv.org/abs/1604.06345}{{\tt
  1604.06345}}].

\bibitem{Durk:2016yja}
J.~Durk and T.~Clifton, \emph{{Exact Initial Data for Black Hole Universes with
  a Cosmological Constant}},
  \href{http://dx.doi.org/10.1088/1361-6382/aa6064}{\emph{Class. Quant. Grav.}
  {\bf 34} (2017) 065009}, [\href{http://arxiv.org/abs/1610.05635}{{\tt
  1610.05635}}].

\bibitem{Bibi:2017urt}
R.~Bibi, T.~Clifton and J.~Durk, \emph{{Cosmological Solutions with Charged
  Black Holes}}, \href{http://dx.doi.org/10.1007/s10714-017-2261-4}{\emph{Gen.
  Rel. Grav.} {\bf 49} (2017) 98}, [\href{http://arxiv.org/abs/1705.01892}{{\tt
  1705.01892}}].

\bibitem{Durk:2017rky}
J.~Durk and T.~Clifton, \emph{{A Quasi-Static Approach to Structure Formation
  in Black Hole Universes}},  \href{http://arxiv.org/abs/1707.08056}{{\tt
  1707.08056}}.

\bibitem{Clifton:2016mxx}
T.~Clifton, D.~Gregoris and K.~Rosquist, \emph{{The Magnetic Part of the Weyl
  Tensor, and the Expansion of Discrete Universes}},
  \href{http://dx.doi.org/10.1007/s10714-017-2192-0}{\emph{Gen. Rel. Grav.}
  {\bf 49} (2017) 30}, [\href{http://arxiv.org/abs/1607.00775}{{\tt
  1607.00775}}].

\bibitem{Wiegand:2010uh}
A.~Wiegand and T.~Buchert, \emph{{Multi-scale cosmology and structure-emerging
  Dark Energy: a plausibility analysis}},
  \href{http://dx.doi.org/10.1103/PhysRevD.82.023523}{\emph{Phys. Rev.} {\bf
  D82} (2010) 023523}, [\href{http://arxiv.org/abs/1002.3912}{{\tt
  1002.3912}}].

\bibitem{Wiegand:2011gs}
A.~Wiegand and T.~Buchert, \emph{{Multiscale approach to inhomogeneous
  cosmologies}}, {\emph{J. Cosmol.} {\bf 15} (2011) 6100},
  [\href{http://arxiv.org/abs/1103.1531}{{\tt 1103.1531}}].

\bibitem{Roukema:2013cya}
B.~F. Roukema, J.~J. Ostrowski and T.~Buchert, \emph{{Virialisation-induced
  curvature as a physical explanation for dark energy}},
  \href{http://dx.doi.org/10.1088/1475-7516/2013/10/043}{\emph{JCAP} {\bf 1310}
  (2013) 043}, [\href{http://arxiv.org/abs/1303.4444}{{\tt 1303.4444}}].

\bibitem{Racz:2016rss}
G.~Racz, L.~Dobos, R.~Beck, I.~Szapudi and I.~Csabai, \emph{{Concordance
  cosmology without dark energy}},
  \href{http://dx.doi.org/10.1093/mnrasl/slx026}{\emph{Mon. Not. Roy. Astron.
  Soc.} {\bf 469} (2017) L1--L5}, [\href{http://arxiv.org/abs/1607.08797}{{\tt
  1607.08797}}].

\bibitem{Roukema:2017doi}
B.~F. Roukema, \emph{{Replacing dark energy by silent virialisation}},
  \href{http://arxiv.org/abs/1706.06179}{{\tt 1706.06179}}.

\bibitem{Adamek:2013wja}
J.~Adamek, D.~Daverio, R.~Durrer and M.~Kunz, \emph{{General Relativistic
  $N$-body simulations in the weak field limit}},
  \href{http://dx.doi.org/10.1103/PhysRevD.88.103527}{\emph{Phys. Rev.} {\bf
  D88} (2013) 103527}, [\href{http://arxiv.org/abs/1308.6524}{{\tt
  1308.6524}}].

\bibitem{Adamek:2014gva}
J.~Adamek, C.~Clarkson, R.~Durrer and M.~Kunz, \emph{{Does small scale
  structure significantly affect cosmological dynamics?}},
  \href{http://dx.doi.org/10.1103/PhysRevLett.114.051302}{\emph{Phys. Rev.
  Lett.} {\bf 114} (2015) 051302}, [\href{http://arxiv.org/abs/1408.2741}{{\tt
  1408.2741}}].

\bibitem{Adamek:2014xba}
J.~Adamek, R.~Durrer and M.~Kunz, \emph{{N-body methods for relativistic
  cosmology}},
  \href{http://dx.doi.org/10.1088/0264-9381/31/23/234006}{\emph{Class. Quant.
  Grav.} {\bf 31} (2014) 234006}, [\href{http://arxiv.org/abs/1408.3352}{{\tt
  1408.3352}}].

\bibitem{Adamek:2015eda}
J.~Adamek, D.~Daverio, R.~Durrer and M.~Kunz, \emph{{General relativity and
  cosmic structure formation}},
  \href{http://dx.doi.org/10.1038/nphys3673}{\emph{Nature Phys.} {\bf 12}
  (2016) 346--349}, [\href{http://arxiv.org/abs/1509.01699}{{\tt 1509.01699}}].

\bibitem{Adamek:2016zes}
J.~Adamek, D.~Daverio, R.~Durrer and M.~Kunz, \emph{{gevolution: a cosmological
  N-body code based on General Relativity}},
  \href{http://dx.doi.org/10.1088/1475-7516/2016/07/053}{\emph{JCAP} {\bf 1607}
  (2016) 053}, [\href{http://arxiv.org/abs/1604.06065}{{\tt 1604.06065}}].

\bibitem{Yoo:2012jz}
C.-M. Yoo, H.~Abe, K.-i. Nakao and Y.~Takamori, \emph{{Black Hole Universe:
  Construction and Analysis of Initial Data}},
  \href{http://dx.doi.org/10.1103/PhysRevD.86.044027}{\emph{Phys. Rev.} {\bf
  D86} (2012) 044027}, [\href{http://arxiv.org/abs/1204.2411}{{\tt
  1204.2411}}].

\bibitem{Bentivegna:2012ei}
E.~Bentivegna and M.~Korzynski, \emph{{Evolution of a periodic eight-black-hole
  lattice in numerical relativity}},
  \href{http://dx.doi.org/10.1088/0264-9381/29/16/165007}{\emph{Class. Quant.
  Grav.} {\bf 29} (2012) 165007}, [\href{http://arxiv.org/abs/1204.3568}{{\tt
  1204.3568}}].

\bibitem{Bentivegna:2013ata}
E.~Bentivegna, \emph{{Black-hole lattices}},
  \href{http://dx.doi.org/10.1007/978-3-642-40157-2_13}{\emph{Springer Proc.
  Math. Stat.} {\bf 60} (2014) 143--146},
  [\href{http://arxiv.org/abs/1307.7673}{{\tt 1307.7673}}].

\bibitem{Adamek:2015hqa}
J.~Adamek, M.~Gosenca and S.~Hotchkiss, \emph{{Spherically Symmetric N-body
  Simulations with General Relativistic Dynamics}},
  \href{http://dx.doi.org/10.1103/PhysRevD.93.023526}{\emph{Phys. Rev.} {\bf
  D93} (2016) 023526}, [\href{http://arxiv.org/abs/1509.01163}{{\tt
  1509.01163}}].

\bibitem{Bentivegna:2016fls}
E.~Bentivegna, M.~Korzyński, I.~Hinder and D.~Gerlicher, \emph{{Light
  propagation through black-hole lattices}},
  \href{http://dx.doi.org/10.1088/1475-7516/2017/03/014}{\emph{JCAP} {\bf 1703}
  (2017) 014}, [\href{http://arxiv.org/abs/1611.09275}{{\tt 1611.09275}}].

\bibitem{bentivegna:2015flc}
E.~Bentivegna and M.~Bruni, \emph{{Effects of nonlinear inhomogeneity on the
  cosmic expansion with numerical relativity}},
  \href{http://dx.doi.org/10.1103/PhysRevLett.116.251302}{\emph{Phys. Rev.
  Lett.} {\bf 116} (2016) 251302}, [\href{http://arxiv.org/abs/1511.05124}{{\tt
  1511.05124}}].

\bibitem{Giblin:2015vwq}
J.~T. Giblin, J.~B. Mertens and G.~D. Starkman, \emph{{Departures from the
  Friedmann-Lemaitre-Robertston-Walker Cosmological Model in an Inhomogeneous
  Universe: A Numerical Examination}},
  \href{http://dx.doi.org/10.1103/PhysRevLett.116.251301}{\emph{Phys. Rev.
  Lett.} {\bf 116} (2016) 251301}, [\href{http://arxiv.org/abs/1511.01105}{{\tt
  1511.01105}}].

\bibitem{Mertens:2015ttp}
J.~B. Mertens, J.~T. Giblin and G.~D. Starkman, \emph{{Integration of
  inhomogeneous cosmological spacetimes in the BSSN formalism}},
  \href{http://dx.doi.org/10.1103/PhysRevD.93.124059}{\emph{Phys. Rev.} {\bf
  D93} (2016) 124059}, [\href{http://arxiv.org/abs/1511.01106}{{\tt
  1511.01106}}].

\bibitem{Macpherson:2016ict}
H.~J. Macpherson, P.~D. Lasky and D.~J. Price, \emph{{Inhomogeneous Cosmology
  with Numerical Relativity}},
  \href{http://dx.doi.org/10.1103/PhysRevD.95.064028}{\emph{Phys. Rev.} {\bf
  D95} (2017) 064028}, [\href{http://arxiv.org/abs/1611.05447}{{\tt
  1611.05447}}].

\bibitem{Giblin:2017juu}
J.~T. Giblin, J.~B. Mertens and G.~D. Starkman, \emph{{A cosmologically
  motivated reference formulation of numerical relativity}},
  \href{http://arxiv.org/abs/1704.04307}{{\tt 1704.04307}}.

\bibitem{Rasanen:2008it}
S.~Rasanen, \emph{{Evaluating backreaction with the peak model of structure
  formation}},
  \href{http://dx.doi.org/10.1088/1475-7516/2008/04/026}{\emph{JCAP} {\bf 0804}
  (2008) 026}, [\href{http://arxiv.org/abs/0801.2692}{{\tt 0801.2692}}].

\bibitem{Shandarin:1984}
S.~F. {Shandarin} and A.~A. {Klypin}, \emph{{Rich Galaxy Clusters may Result
  from Largescale Motions Inside Superclusters}}, {\emph{\sovast} {\bf 28}
  (Oct., 1984) 491--495}.

\bibitem{Katz:1993}
N.~{Katz}, T.~{Quinn} and J.~M. {Gelb}, \emph{{Galaxy Formation and the Peaks
  Formalism}}, \href{http://dx.doi.org/10.1093/mnras/265.3.689}{\emph{\mnras}
  {\bf 265} (Dec., 1993) 689}.

\bibitem{vandeWeygaert:1994ww}
R.~van~de Weygaert and A.~Babul, \emph{{Shear fields and the evolution of
  galactic scale density peaks}},
  \href{http://dx.doi.org/10.1086/187310}{\emph{Astrophys. J.} {\bf 425} (1994)
  L59}, [\href{http://arxiv.org/abs/astro-ph/9402003}{{\tt astro-ph/9402003}}].

\bibitem{Kofman:1994pz}
L.~Kofman and D.~Pogosian, \emph{{Equations of gravitational instability are
  nonlocal}}, \href{http://dx.doi.org/10.1086/175419}{\emph{Astrophys. J.} {\bf
  442} (1995) 30--38}, [\href{http://arxiv.org/abs/astro-ph/9403029}{{\tt
  astro-ph/9403029}}].

\bibitem{Porciani:2001er}
C.~Porciani, A.~Dekel and Y.~Hoffman, \emph{{Testing tidal-torque theory. 2.
  Alignment of inertia and shear and the characteristics of proto-haloes}},
  \href{http://dx.doi.org/10.1046/j.1365-8711.2002.05306.x}{\emph{Mon. Not.
  Roy. Astron. Soc.} {\bf 332} (2002) 339},
  [\href{http://arxiv.org/abs/astro-ph/0105165}{{\tt astro-ph/0105165}}].

\bibitem{Ludlow:2010xd}
A.~D. Ludlow and C.~Porciani, \emph{{The Peaks Formalism and the Formation of
  Cold Dark Matter Haloes}},
  \href{http://dx.doi.org/10.1111/j.1365-2966.2011.18282.x}{\emph{Mon. Not.
  Roy. Astron. Soc.} {\bf 413} (2011) 1961--1972},
  [\href{http://arxiv.org/abs/1011.2493}{{\tt 1011.2493}}].

\bibitem{Bond:1995yt}
J.~R. Bond, L.~Kofman and D.~Pogosyan, \emph{{How filaments are woven into the
  cosmic web}}, \href{http://dx.doi.org/10.1038/380603a0}{\emph{Nature} {\bf
  380} (1996) 603--606}, [\href{http://arxiv.org/abs/astro-ph/9512141}{{\tt
  astro-ph/9512141}}].

\bibitem{Cautun:2014fwa}
M.~Cautun, R.~van~de Weygaert, B.~J.~T. Jones and C.~S. Frenk, \emph{{Evolution
  of the cosmic web}}, \href{http://dx.doi.org/10.1093/mnras/stu768}{\emph{Mon.
  Not. Roy. Astron. Soc.} {\bf 441} (2014) 2923--2973},
  [\href{http://arxiv.org/abs/1401.7866}{{\tt 1401.7866}}].

\bibitem{Zeldovich:1970}
Y.~B. {Zel'dovich}, \emph{{Gravitational instability: An approximate theory for
  large density perturbations.}}, {\emph{\aap} {\bf 5} (Mar., 1970) 84--89}.

\bibitem{Icke:1973}
V.~{Icke}, \emph{{Formation of Galaxies inside Clusters}}, {\emph{\aap} {\bf
  27} (Aug., 1973) 1}.

\bibitem{White:1979}
S.~D.~M. {White} and J.~{Silk}, \emph{{The growth of aspherical structure in
  the universe - Is the local supercluster an unusual system}},
  \href{http://dx.doi.org/10.1086/157156}{\emph{\apj} {\bf 231} (July, 1979)
  1--9}.

\bibitem{Peebles:1980}
P.~J.~E. {Peebles} and G.~{Shaviv}, \emph{{The Large-Scale Structure of the
  Universe}}, {\emph{\ssr} {\bf 31} (Mar., 1982) 119}.

\bibitem{Hoffman:1986}
Y.~{Hoffman}, \emph{{The dynamics of superclusters - The effect of shear}},
  \href{http://dx.doi.org/10.1086/164520}{\emph{\apj} {\bf 308} (Sept., 1986)
  493--498}.

\bibitem{Eisenstein:1994ni}
D.~J. Eisenstein and A.~Loeb, \emph{{An Analytical model for the triaxial
  collapse of cosmological perturbations}},
  \href{http://dx.doi.org/10.1086/175193}{\emph{Astrophys. J.} {\bf 439} (1995)
  520}, [\href{http://arxiv.org/abs/astro-ph/9405012}{{\tt astro-ph/9405012}}].

\bibitem{vandeWeygaert:1995pz}
R.~van~de Weygaert and E.~Bertschinger, \emph{{Constraining peaks in Gaussian
  primordial density fields: an application of the hoffman-ribak method}},
  \href{http://dx.doi.org/10.1093/mnras/281.1.84}{\emph{Mon. Not. Roy. Astron.
  Soc.} {\bf 281} (1996) 84},
  [\href{http://arxiv.org/abs/astro-ph/9507024}{{\tt astro-ph/9507024}}].

\bibitem{Audit:1996zj}
E.~Audit and J.-M. Alimi, \emph{{Gravitational Lagrangian dynamics of cold
  matter using the deformation tensor}}, {\emph{Astron. Astrophys.} {\bf 315}
  (1996) 11--20}, [\href{http://arxiv.org/abs/astro-ph/9609156}{{\tt
  astro-ph/9609156}}].

\bibitem{Bond:1996}
J.~R. {Bond} and S.~T. {Myers}, \emph{{The Peak-Patch Picture of Cosmic
  Catalogs. I. Algorithms}},
  \href{http://dx.doi.org/10.1086/192267}{\emph{\apjs} {\bf 103} (Mar., 1996)
  1}.

\bibitem{DelPopolo:2001fq}
A.~Del~Popolo, E.~N. Ercan and Z.~Xia, \emph{{Ellipsoidal collapse and
  previrialization}}, \href{http://dx.doi.org/10.1086/321137}{\emph{Astron. J.}
  {\bf 122} (2001) 487--495},
  [\href{http://arxiv.org/abs/astro-ph/0108080}{{\tt astro-ph/0108080}}].

\bibitem{Popolo:2002}
A.~D. Popolo, \emph{{On the evolution of aspherical perturbations in the
  universe: an analytical model}},
  \href{http://dx.doi.org/10.1051/0004-6361:20020399}{\emph{Astron. Astrophys.}
  {\bf 387} (2002) 759}, [\href{http://arxiv.org/abs/astro-ph/0202436}{{\tt
  astro-ph/0202436}}].

\bibitem{Sheth:1999su}
R.~K. Sheth, H.~J. Mo and G.~Tormen, \emph{{Ellipsoidal collapse and an
  improved model for the number and spatial distribution of dark matter
  haloes}},
  \href{http://dx.doi.org/10.1046/j.1365-8711.2001.04006.x}{\emph{Mon. Not.
  Roy. Astron. Soc.} {\bf 323} (2001) 1},
  [\href{http://arxiv.org/abs/astro-ph/9907024}{{\tt astro-ph/9907024}}].

\bibitem{Sheth:2001dp}
R.~K. Sheth and G.~Tormen, \emph{{An Excursion set model of hierarchical
  clustering : Ellipsoidal collapse and the moving barrier}},
  \href{http://dx.doi.org/10.1046/j.1365-8711.2002.04950.x}{\emph{Mon. Not.
  Roy. Astron. Soc.} {\bf 329} (2002) 61},
  [\href{http://arxiv.org/abs/astro-ph/0105113}{{\tt astro-ph/0105113}}].

\bibitem{Ohta:2003zc}
Y.~Ohta, I.~Kayo and A.~Taruya, \emph{{Evolution of cosmological density
  distribution function from the local collapse model}},
  \href{http://dx.doi.org/10.1086/374375}{\emph{Astrophys. J.} {\bf 589} (2003)
  1--16}, [\href{http://arxiv.org/abs/astro-ph/0301567}{{\tt
  astro-ph/0301567}}].

\bibitem{Ohta:2004mx}
Y.~Ohta, I.~Kayo and A.~Taruya, \emph{{Cosmological density distribution
  function from the ellipsoidal collapse model in real space}},
  \href{http://dx.doi.org/10.1086/420762}{\emph{Astrophys. J.} {\bf 608} (2004)
  647--662}, [\href{http://arxiv.org/abs/astro-ph/0402618}{{\tt
  astro-ph/0402618}}].

\bibitem{Lam:2008ik}
T.~Y. Lam and R.~K. Sheth, \emph{{Ellipsoidal collapse and the redshift space
  probability distribution function of dark matter}},
  \href{http://dx.doi.org/10.1111/j.1365-2966.2008.13621.x}{\emph{Mon. Not.
  Roy. Astron. Soc.} {\bf 389} (2008) 1249},
  [\href{http://arxiv.org/abs/0805.1238}{{\tt 0805.1238}}].

\bibitem{Robertson:2008jr}
B.~E. Robertson, A.~V. Kravtsov, J.~Tinker and A.~R. Zentner, \emph{{Collapse
  Barriers and Halo Abundance: Testing the Excursion Set Ansatz}},
  \href{http://dx.doi.org/10.1088/0004-637X/696/1/636}{\emph{Astrophys. J.}
  {\bf 696} (2009) 636--652}, [\href{http://arxiv.org/abs/0812.3148}{{\tt
  0812.3148}}].

\bibitem{Angrick:2010qg}
C.~Angrick and M.~Bartelmann, \emph{{Triaxial collapse and virialisation of
  dark-matter haloes}},
  \href{http://dx.doi.org/10.1051/0004-6361/201014147}{\emph{Astron.
  Astrophys.} {\bf 518} (2010) A38},
  [\href{http://arxiv.org/abs/1001.4984}{{\tt 1001.4984}}].

\bibitem{Ludlow:2011jx}
A.~D. Ludlow, C.~Porciani and M.~Borzyszkowski, \emph{{The formation of CDM
  haloes – I. Collapse thresholds and the ellipsoidal collapse model}},
  \href{http://dx.doi.org/10.1093/mnras/stu2021}{\emph{Mon. Not. Roy. Astron.
  Soc.} {\bf 445} (2014) 4110--4123},
  [\href{http://arxiv.org/abs/1107.5808}{{\tt 1107.5808}}].

\bibitem{Paranjape:2012ks}
A.~Paranjape and R.~K. Sheth, \emph{{Peaks theory and the excursion set
  approach}},
  \href{http://dx.doi.org/10.1111/j.1365-2966.2012.21911.x}{\emph{Mon. Not.
  Roy. Astron. Soc.} {\bf 426} (2012) 2789--2796},
  [\href{http://arxiv.org/abs/1206.3506}{{\tt 1206.3506}}].

\bibitem{Paranjape:2012jt}
A.~Paranjape, R.~K. Sheth and V.~Desjacques, \emph{{Excursion set peaks: a
  self-consistent model of dark halo abundances and clustering}},
  \href{http://dx.doi.org/10.1093/mnras/stt267}{\emph{Mon. Not. Roy. Astron.
  Soc.} {\bf 431} (2013) 1503--1512},
  [\href{http://arxiv.org/abs/1210.1483}{{\tt 1210.1483}}].

\bibitem{Reischke:2016dop}
R.~Reischke, F.~Pace, S.~Meyer and B.~M. Schäfer, \emph{{Spherical collapse of
  dark matter haloes in tidal gravitational fields}},
  \href{http://dx.doi.org/10.1093/mnras/stw1989}{\emph{Mon. Not. Roy. Astron.
  Soc.} {\bf 463} (2016) 429--440},
  [\href{http://arxiv.org/abs/1606.09207}{{\tt 1606.09207}}].

\bibitem{Buchert:1999pq}
T.~Buchert, M.~Kerscher and C.~Sicka, \emph{{Back reaction of inhomogeneities
  on the expansion: The Evolution of cosmological parameters}},
  \href{http://dx.doi.org/10.1103/PhysRevD.62.043525}{\emph{Phys. Rev.} {\bf
  D62} (2000) 043525}, [\href{http://arxiv.org/abs/astro-ph/9912347}{{\tt
  astro-ph/9912347}}].

\bibitem{Chuang:2005yi}
C.-H. Chuang, J.-A. Gu and W.-Y.~P. Hwang, \emph{{Inhomogeneity-induced cosmic
  acceleration in a dust universe}},
  \href{http://dx.doi.org/10.1088/0264-9381/25/17/175001}{\emph{Class. Quant.
  Grav.} {\bf 25} (2008) 175001},
  [\href{http://arxiv.org/abs/astro-ph/0512651}{{\tt astro-ph/0512651}}].

\bibitem{Paranjape:2006cd}
A.~Paranjape and T.~P. Singh, \emph{{The Possibility of Cosmic Acceleration via
  Spatial Averaging in Lemaitre-Tolman-Bondi Models}},
  \href{http://dx.doi.org/10.1088/0264-9381/23/23/022}{\emph{Class. Quant.
  Grav.} {\bf 23} (2006) 6955--6969},
  [\href{http://arxiv.org/abs/astro-ph/0605195}{{\tt astro-ph/0605195}}].

\bibitem{Kai:2006ws}
T.~Kai, H.~Kozaki, K.-i. nakao, Y.~Nambu and C.-M. Yoo, \emph{{Can
  inhomogeneties accelerate the cosmic volume expansion?}},
  \href{http://dx.doi.org/10.1143/PTP.117.229}{\emph{Prog. Theor. Phys.} {\bf
  117} (2007) 229--240}, [\href{http://arxiv.org/abs/gr-qc/0605120}{{\tt
  gr-qc/0605120}}].

\bibitem{Buchert:1999er}
T.~Buchert, \emph{{On average properties of inhomogeneous fluids in general
  relativity. 1. Dust cosmologies}},
  \href{http://dx.doi.org/10.1023/A:1001800617177}{\emph{Gen. Rel. Grav.} {\bf
  32} (2000) 105--125}, [\href{http://arxiv.org/abs/gr-qc/9906015}{{\tt
  gr-qc/9906015}}].

\bibitem{Buchert:2001sa}
T.~Buchert, \emph{{On average properties of inhomogeneous fluids in general
  relativity: Perfect fluid cosmologies}},
  \href{http://dx.doi.org/10.1023/A:1012061725841}{\emph{Gen. Rel. Grav.} {\bf
  33} (2001) 1381--1405}, [\href{http://arxiv.org/abs/gr-qc/0102049}{{\tt
  gr-qc/0102049}}].

\bibitem{Larena:2009md}
J.~Larena, \emph{{Spatially averaged cosmology in an arbitrary coordinate
  system}}, \href{http://dx.doi.org/10.1103/PhysRevD.79.084006}{\emph{Phys.
  Rev.} {\bf D79} (2009) 084006}, [\href{http://arxiv.org/abs/0902.3159}{{\tt
  0902.3159}}].

\bibitem{Malament:1995}
D.~D. {Malament}, \emph{{Is Newtonian cosmology really inconsistent?}},
  {\emph{Philosophy of Science} {\bf 62} (1995) 489}.

\bibitem{Norton:1995}
J.~D. {Norton}, \emph{{The force of Newtonian cosmology: acceleration is
  relative}}, {\emph{Philosophy of Science} {\bf 62} (1995) 511}.

\bibitem{Ade:2015xua}
{\scshape Planck} collaboration, P.~A.~R. Ade et~al., \emph{{Planck 2015
  results. XIII. Cosmological parameters}},
  \href{http://dx.doi.org/10.1051/0004-6361/201525830}{\emph{Astron.
  Astrophys.} {\bf 594} (2016) A13},
  [\href{http://arxiv.org/abs/1502.01589}{{\tt 1502.01589}}].

\bibitem{Bardeen:1986}
J.~M. {Bardeen}, J.~R. {Bond}, N.~{Kaiser} and A.~S. {Szalay}, \emph{{The
  statistics of peaks of Gaussian random fields}},
  \href{http://dx.doi.org/10.1086/164143}{\emph{\apj} {\bf 304} (May, 1986)
  15--61}.

\bibitem{Doroshkevich:1970}
A.~G. {Doroshkevich}, \emph{{The space structure of perturbations and the
  origin of rotation of galaxies in the theory of fluctuation.}},
  {\emph{Astrofizika} {\bf 6} (1970) 581--600}.

\bibitem{Angrick:2013wba}
C.~Angrick, \emph{{Ellipticity and prolaticity of the initial
  gravitational-shear field at the position of density maxima}},
  \href{http://dx.doi.org/10.1093/mnras/stu1272}{\emph{Mon. Not. Roy. Astron.
  Soc.} {\bf 443} (2014) 2361--2371},
  [\href{http://arxiv.org/abs/1305.0497}{{\tt 1305.0497}}].

\bibitem{Lewis:1999bs}
A.~Lewis, A.~Challinor and A.~Lasenby, \emph{{Efficient computation of CMB
  anisotropies in closed FRW models}},
  \href{http://dx.doi.org/10.1086/309179}{\emph{Astrophys. J.} {\bf 538} (2000)
  473--476}, [\href{http://arxiv.org/abs/astro-ph/9911177}{{\tt
  astro-ph/9911177}}].

\bibitem{Vonlanthen:2010cd}
M.~Vonlanthen, S.~Rasanen and R.~Durrer, \emph{{Model-independent cosmological
  constraints from the CMB}},
  \href{http://dx.doi.org/10.1088/1475-7516/2010/08/023}{\emph{JCAP} {\bf 1008}
  (2010) 023}, [\href{http://arxiv.org/abs/1003.0810}{{\tt 1003.0810}}].

\bibitem{Audren:2012wb}
B.~Audren, J.~Lesgourgues, K.~Benabed and S.~Prunet, \emph{{Conservative
  Constraints on Early Cosmology: an illustration of the Monte Python
  cosmological parameter inference code}},
  \href{http://dx.doi.org/10.1088/1475-7516/2013/02/001}{\emph{JCAP} {\bf 1302}
  (2013) 001}, [\href{http://arxiv.org/abs/1210.7183}{{\tt 1210.7183}}].

\bibitem{Audren:2013nwa}
B.~Audren, \emph{{Separate Constraints on Early and Late Cosmology}},
  \href{http://dx.doi.org/10.1093/mnras/stu1457}{\emph{Mon. Not. Roy. Astron.
  Soc.} {\bf 444} (2014) 827--832}, [\href{http://arxiv.org/abs/1312.5696}{{\tt
  1312.5696}}].

\bibitem{Rasanen:2006zw}
S.~Rasanen, \emph{{Cosmological acceleration from structure formation}},
  \href{http://dx.doi.org/10.1142/S0218271806009728}{\emph{Int. J. Mod. Phys.}
  {\bf D15} (2006) 2141--2146},
  [\href{http://arxiv.org/abs/astro-ph/0605632}{{\tt astro-ph/0605632}}].

\bibitem{Rasanen:2006kp}
S.~Rasanen, \emph{{Accelerated expansion from structure formation}},
  \href{http://dx.doi.org/10.1088/1475-7516/2006/11/003}{\emph{JCAP} {\bf 0611}
  (2006) 003}, [\href{http://arxiv.org/abs/astro-ph/0607626}{{\tt
  astro-ph/0607626}}].

\bibitem{Shapiro:2005nz}
C.~Shapiro and M.~S. Turner, \emph{{What do we really know about cosmic
  acceleration?}}, \href{http://dx.doi.org/10.1086/506470}{\emph{Astrophys. J.}
  {\bf 649} (2006) 563--569},
  [\href{http://arxiv.org/abs/astro-ph/0512586}{{\tt astro-ph/0512586}}].

\bibitem{Gong:2006tx}
Y.~Gong and A.~Wang, \emph{{Observational constraints on the acceleration of
  the universe}},
  \href{http://dx.doi.org/10.1103/PhysRevD.73.083506}{\emph{Phys. Rev.} {\bf
  D73} (2006) 083506}, [\href{http://arxiv.org/abs/astro-ph/0601453}{{\tt
  astro-ph/0601453}}].

\bibitem{Elgaroy:2006tp}
O.~Elgaroy and T.~Multamaki, \emph{{Bayesian analysis of friedmannless
  cosmologies}},
  \href{http://dx.doi.org/10.1088/1475-7516/2006/09/002}{\emph{JCAP} {\bf 0609}
  (2006) 002}, [\href{http://arxiv.org/abs/astro-ph/0603053}{{\tt
  astro-ph/0603053}}].

\bibitem{Seikel:2007pk}
M.~Seikel and D.~J. Schwarz, \emph{{How strong is the evidence for accelerated
  expansion?}},
  \href{http://dx.doi.org/10.1088/1475-7516/2008/02/007}{\emph{JCAP} {\bf 0802}
  (2008) 007}, [\href{http://arxiv.org/abs/0711.3180}{{\tt 0711.3180}}].

\bibitem{Seikel:2008ms}
M.~Seikel and D.~J. Schwarz, \emph{{Model- and calibration-independent test of
  cosmic acceleration}},
  \href{http://dx.doi.org/10.1088/1475-7516/2009/02/024}{\emph{JCAP} {\bf 0902}
  (2009) 024}, [\href{http://arxiv.org/abs/0810.4484}{{\tt 0810.4484}}].

\bibitem{Mortsell:2008yu}
E.~Mortsell and C.~Clarkson, \emph{{Model independent constraints on the
  cosmological expansion rate}},
  \href{http://dx.doi.org/10.1088/1475-7516/2009/01/044}{\emph{JCAP} {\bf 0901}
  (2009) 044}, [\href{http://arxiv.org/abs/0811.0981}{{\tt 0811.0981}}].

\bibitem{Guimaraes:2009mp}
A.~C.~C. Guimaraes, J.~V. Cunha and J.~A.~S. Lima, \emph{{Bayesian Analysis and
  Constraints on Kinematic Models from Union SNIa}},
  \href{http://dx.doi.org/10.1088/1475-7516/2009/10/010}{\emph{JCAP} {\bf 0910}
  (2009) 010}, [\href{http://arxiv.org/abs/0904.3550}{{\tt 0904.3550}}].

\bibitem{Serra:2009yp}
P.~Serra, A.~Cooray, D.~E. Holz, A.~Melchiorri, S.~Pandolfi and D.~Sarkar,
  \emph{{No Evidence for Dark Energy Dynamics from a Global Analysis of
  Cosmological Data}},
  \href{http://dx.doi.org/10.1103/PhysRevD.80.121302}{\emph{Phys. Rev.} {\bf
  D80} (2009) 121302}, [\href{http://arxiv.org/abs/0908.3186}{{\tt
  0908.3186}}].

\bibitem{Cai:2010qp}
R.-G. Cai, Q.~Su and H.-B. Zhang, \emph{{Probing the dynamical behavior of dark
  energy}}, \href{http://dx.doi.org/10.1088/1475-7516/2010/04/012}{\emph{JCAP}
  {\bf 1004} (2010) 012}, [\href{http://arxiv.org/abs/1001.2207}{{\tt
  1001.2207}}].

\bibitem{Wang:2010vj}
S.~Wang, X.-D. Li and M.~Li, \emph{{Exploring the Latest Union2 SNIa Dataset by
  Using Model-Independent Parametrization Methods}},
  \href{http://dx.doi.org/10.1103/PhysRevD.83.023010}{\emph{Phys. Rev.} {\bf
  D83} (2011) 023010}, [\href{http://arxiv.org/abs/1009.5837}{{\tt
  1009.5837}}].

\bibitem{Park:2010xw}
J.~Park, C.-G. Park and J.-c. Hwang, \emph{{Analysis of recent type Ia
  supernova data based on evolving dark energy models}},
  \href{http://dx.doi.org/10.1103/PhysRevD.84.023506}{\emph{Phys. Rev.} {\bf
  D84} (2011) 023506}, [\href{http://arxiv.org/abs/1011.1723}{{\tt
  1011.1723}}].

\bibitem{Pan:2010zh}
A.~V. Pan and U.~Alam, \emph{{Reconstructing Dark Energy : A Comparison of
  Cosmological Parameters}},  \href{http://arxiv.org/abs/1012.1591}{{\tt
  1012.1591}}.

\bibitem{Cai:2011px}
R.-G. Cai and Z.-L. Tuo, \emph{{Detecting the cosmic acceleration with current
  data}}, \href{http://dx.doi.org/10.1016/j.physletb.2011.11.021}{\emph{Phys.
  Lett.} {\bf B706} (2011) 116--122},
  [\href{http://arxiv.org/abs/1105.1603}{{\tt 1105.1603}}].

\bibitem{Li:2011wb}
X.-D. Li, S.~Li, S.~Wang, W.-S. Zhang, Q.-G. Huang and M.~Li, \emph{{Probing
  Cosmic Acceleration by Using the SNLS3 SNIa Dataset}},
  \href{http://dx.doi.org/10.1088/1475-7516/2011/07/011}{\emph{JCAP} {\bf 1107}
  (2011) 011}, [\href{http://arxiv.org/abs/1106.4116}{{\tt 1106.4116}}].

\bibitem{Shafieloo:2012ht}
A.~Shafieloo, A.~G. Kim and E.~V. Linder, \emph{{Gaussian Process
  Cosmography}},
  \href{http://dx.doi.org/10.1103/PhysRevD.85.123530}{\emph{Phys. Rev.} {\bf
  D85} (2012) 123530}, [\href{http://arxiv.org/abs/1204.2272}{{\tt
  1204.2272}}].

\bibitem{Aviles:2012ay}
A.~Aviles, C.~Gruber, O.~Luongo and H.~Quevedo, \emph{{Cosmography and
  constraints on the equation of state of the Universe in various
  parametrizations}},
  \href{http://dx.doi.org/10.1103/PhysRevD.86.123516}{\emph{Phys. Rev.} {\bf
  D86} (2012) 123516}, [\href{http://arxiv.org/abs/1204.2007}{{\tt
  1204.2007}}].

\bibitem{Magana:2014voa}
J.~Magaña, V.~H. Cárdenas and V.~Motta, \emph{{Cosmic slowing down of
  acceleration for several dark energy parametrizations}},
  \href{http://dx.doi.org/10.1088/1475-7516/2014/10/017}{\emph{JCAP} {\bf 1410}
  (2014) 017}, [\href{http://arxiv.org/abs/1407.1632}{{\tt 1407.1632}}].

\bibitem{Qing-Guo:2016ykt}
Q.-G. Huang and K.~Wang, \emph{{How the dark energy can reconcile Planck with
  local determination of the Hubble constant}},
  \href{http://dx.doi.org/10.1140/epjc/s10052-016-4352-x}{\emph{Eur. Phys. J.}
  {\bf C76} (2016) 506}, [\href{http://arxiv.org/abs/1606.05965}{{\tt
  1606.05965}}].

\bibitem{Ade:2015rim}
{\scshape Planck} collaboration, P.~A.~R. Ade et~al., \emph{{Planck 2015
  results. XIV. Dark energy and modified gravity}},
  \href{http://dx.doi.org/10.1051/0004-6361/201525814}{\emph{Astron.
  Astrophys.} {\bf 594} (2016) A14},
  [\href{http://arxiv.org/abs/1502.01590}{{\tt 1502.01590}}].

\bibitem{Cattoen:2007id}
C.~Cattoen and M.~Visser, \emph{{Cosmography: Extracting the Hubble series from
  the supernova data}},  \href{http://arxiv.org/abs/gr-qc/0703122}{{\tt
  gr-qc/0703122}}.

\bibitem{Visser:2009zs}
M.~Visser and C.~Cattoen, \emph{{Cosmographic analysis of dark energy}},  in
  \emph{{Proceedings, 7th International Heidelberg Conference on Dark Matter in
  Astro and Particle Physics (DARK 2009): Christchurch, New Zealand, January
  18-24, 2009}}, pp.~287--300, 2009.
\newblock \href{http://arxiv.org/abs/0906.5407}{{\tt 0906.5407}}.
\newblock \href{http://dx.doi.org/10.1142/9789814293792_0022}{DOI}.

\bibitem{Clarkson:2011gm}
C.~Clarkson, T.~Clifton, A.~Coley and R.~Sung, \emph{{Observational Constraints
  on the Averaged Universe}},
  \href{http://dx.doi.org/10.1103/PhysRevD.90.049903,
  10.1103/PhysRevD.85.043506}{\emph{Phys. Rev.} {\bf D85} (2012) 043506},
  [\href{http://arxiv.org/abs/1111.2214}{{\tt 1111.2214}}].

\bibitem{Okouma:2012dy}
P.~M. Okouma, Y.~Fantaye and B.~A. Bassett, \emph{{How Flat is Our Universe
  Really?}},
  \href{http://dx.doi.org/10.1016/j.physletb.2012.12.070}{\emph{Phys. Lett.}
  {\bf B719} (2013) 1--4}, [\href{http://arxiv.org/abs/1207.3000}{{\tt
  1207.3000}}].

\bibitem{Zhang:2017aqn}
B.~R. Zhang, M.~J. Childress, T.~M. Davis, N.~V. Karpenka, C.~Lidman, B.~P.
  Schmidt et~al., \emph{{A blinded determination of $H_0$ from low-redshift
  Type Ia supernovae, calibrated by Cepheid variables}},
  \href{http://arxiv.org/abs/1706.07573}{{\tt 1706.07573}}.

\bibitem{Efstathiou:2013via}
G.~Efstathiou, \emph{{H0 Revisited}},
  \href{http://dx.doi.org/10.1093/mnras/stu278}{\emph{Mon. Not. Roy. Astron.
  Soc.} {\bf 440} (2014) 1138--1152},
  [\href{http://arxiv.org/abs/1311.3461}{{\tt 1311.3461}}].

\bibitem{Riess:2016jrr}
A.~G. Riess et~al., \emph{{A 2.4\% Determination of the Local Value of the
  Hubble Constant}},
  \href{http://dx.doi.org/10.3847/0004-637X/826/1/56}{\emph{Astrophys. J.} {\bf
  826} (2016) 56}, [\href{http://arxiv.org/abs/1604.01424}{{\tt 1604.01424}}].

\bibitem{Cardona:2016ems}
W.~Cardona, M.~Kunz and V.~Pettorino, \emph{{Determining $H_0$ with Bayesian
  hyper-parameters}},
  \href{http://dx.doi.org/10.1088/1475-7516/2017/03/056}{\emph{JCAP} {\bf 1703}
  (2017) 056}, [\href{http://arxiv.org/abs/1611.06088}{{\tt 1611.06088}}].

\bibitem{Follin:2017ljs}
B.~Follin and L.~Knox, \emph{{Insensitivity of The Distance Ladder Hubble
  Constant Determination to Cepheid Calibration Modeling Choices}},
  \href{http://arxiv.org/abs/1707.01175}{{\tt 1707.01175}}.

\bibitem{Feeney:2017sgx}
S.~M. Feeney, D.~J. Mortlock and N.~Dalmasso, \emph{{Clarifying the Hubble
  constant tension with a Bayesian hierarchical model of the local distance
  ladder}},  \href{http://arxiv.org/abs/1707.00007}{{\tt 1707.00007}}.

\bibitem{Krauss:2003}
L.~M. {Krauss} and B.~{Chaboyer}, \emph{{Age Estimates of Globular Clusters in
  the Milky Way: Constraints on Cosmology}},
  \href{http://dx.doi.org/10.1126/science.1075631}{\emph{Science} {\bf 299}
  (Jan., 2003) 65--70}.

\bibitem{Hofmann:2002nu}
S.~Hofmann, D.~J. Schwarz and H.~Stoecker, \emph{{Formation of small-scale
  structure in SUSY CDM}},  in \emph{{Proceedings, 4th International Workshop
  on The identification of dark matter (IDM 2002): York, UK, September 2-6,
  2002}}, pp.~45--51, 2002.
\newblock \href{http://arxiv.org/abs/astro-ph/0211325}{{\tt astro-ph/0211325}}.
\newblock \href{http://dx.doi.org/10.1142/9789812791313_0006}{DOI}.

\bibitem{Green:2003un}
A.~M. Green, S.~Hofmann and D.~J. Schwarz, \emph{{The power spectrum of SUSY -
  CDM on sub-galactic scales}},
  \href{http://dx.doi.org/10.1111/j.1365-2966.2004.08232.x}{\emph{Mon. Not.
  Roy. Astron. Soc.} {\bf 353} (2004) L23},
  [\href{http://arxiv.org/abs/astro-ph/0309621}{{\tt astro-ph/0309621}}].

\bibitem{Green:2005fa}
A.~M. Green, S.~Hofmann and D.~J. Schwarz, \emph{{The First wimpy halos}},
  \href{http://dx.doi.org/10.1088/1475-7516/2005/08/003}{\emph{JCAP} {\bf 0508}
  (2005) 003}, [\href{http://arxiv.org/abs/astro-ph/0503387}{{\tt
  astro-ph/0503387}}].

\bibitem{Green:2005kf}
A.~M. Green, S.~Hofmann and D.~J. Schwarz, \emph{{Small scale wimp physics}},
  \href{http://dx.doi.org/10.1063/1.2149748}{\emph{AIP Conf. Proc.} {\bf 805}
  (2006) 431--434}, [\href{http://arxiv.org/abs/astro-ph/0508553}{{\tt
  astro-ph/0508553}}].

\bibitem{Szekeres:1975}
P.~{Szekeres}, \emph{{A class of inhomogeneous cosmological models}},
  \href{http://dx.doi.org/10.1007/BF01608547}{\emph{Communications in
  Mathematical Physics} {\bf 41} (Feb., 1975) 55--64}.

\bibitem{Reischke:2016eza}
R.~Reischke, F.~Pace, S.~Meyer and B.~M. Schafer, \emph{{Shear and vorticity in
  the spherical collapse of dark matter haloes}},
  \href{http://arxiv.org/abs/1612.04275}{{\tt 1612.04275}}.

\bibitem{vanElst:1996zs}
H.~van Elst, C.~Uggla, W.~M. Lesame, G.~F.~R. Ellis and R.~Maartens,
  \emph{{Integrability of irrotational silent cosmological models}},
  \href{http://dx.doi.org/10.1088/0264-9381/14/5/018}{\emph{Class. Quant.
  Grav.} {\bf 14} (1997) 1151--1162},
  [\href{http://arxiv.org/abs/gr-qc/9611002}{{\tt gr-qc/9611002}}].

\bibitem{Sopuerta:1997}
C.~F. {Sopuerta}, \emph{{New study of silent universes}},
  \href{http://dx.doi.org/10.1103/PhysRevD.55.5936}{\emph{\prd} {\bf 55} (May,
  1997) 5936--5950}.

\bibitem{Ellis:1971pg}
G.~F.~R. Ellis, \emph{{Relativistic cosmology}},
  \href{http://dx.doi.org/10.1007/s10714-009-0760-7}{\emph{Gen. Rel. Grav.}
  {\bf 41} (2009) 581--660}.

\bibitem{Ellis:1994md}
G.~F.~R. Ellis and P.~K.~S. Dunsby, \emph{{Newtonian evolution of the Weyl
  tensor}}, \href{http://dx.doi.org/10.1086/303839}{\emph{Astrophys. J.} {\bf
  479} (1997) 97}, [\href{http://arxiv.org/abs/astro-ph/9410001}{{\tt
  astro-ph/9410001}}].

\bibitem{Matarrese:1995sb}
S.~Matarrese and D.~Terranova, \emph{{PostNewtonian cosmological dynamics in
  Lagrangian coordinates}},
  \href{http://dx.doi.org/10.1093/mnras/283.2.400}{\emph{Mon. Not. Roy. Astron.
  Soc.} {\bf 283} (1996) 400--418},
  [\href{http://arxiv.org/abs/astro-ph/9511093}{{\tt astro-ph/9511093}}].

\bibitem{vanElst:1998kb}
H.~van Elst and G.~F.~R. Ellis, \emph{{QuasiNewtonian dust cosmologies}},
  \href{http://dx.doi.org/10.1088/0264-9381/15/11/017}{\emph{Class. Quant.
  Grav.} {\bf 15} (1998) 3545--3573},
  [\href{http://arxiv.org/abs/gr-qc/9805087}{{\tt gr-qc/9805087}}].

\bibitem{Ehlers:1997}
J.~{Ehlers}, \emph{{Examples of Newtonian limits of relativistic spacetimes}},
  \href{http://dx.doi.org/10.1088/0264-9381/14/1A/010}{\emph{Classical and
  Quantum Gravity} {\bf 14} (Jan., 1997) A119--A126}.

\bibitem{Ehlers:2009uv}
J.~Ehlers and T.~Buchert, \emph{{On the Newtonian Limit of the Weyl Tensor}},
  \href{http://dx.doi.org/10.1007/s10714-009-0855-1}{\emph{Gen. Rel. Grav.}
  {\bf 41} (2009) 2153--2158}, [\href{http://arxiv.org/abs/0907.2645}{{\tt
  0907.2645}}].

\bibitem{Rasanen:2011bm}
S.~Rasanen, \emph{{Light propagation and the average expansion rate in near-FRW
  universes}}, \href{http://dx.doi.org/10.1103/PhysRevD.85.083528}{\emph{Phys.
  Rev.} {\bf D85} (2012) 083528}, [\href{http://arxiv.org/abs/1107.1176}{{\tt
  1107.1176}}].

\bibitem{Buchert:2017obp}
T.~Buchert, \emph{{Comment on: "Why there is no Newtonian backreaction" by N.
  Kaiser}},  \href{http://arxiv.org/abs/1704.00703}{{\tt 1704.00703}}.

\bibitem{Gibbons:2013msa}
G.~F.~R. Ellis and G.~W. Gibbons, \emph{{Discrete Newtonian Cosmology}},
  \href{http://dx.doi.org/10.1088/0264-9381/31/2/025003}{\emph{Class. Quant.
  Grav.} {\bf 31} (2014) 025003}, [\href{http://arxiv.org/abs/1308.1852}{{\tt
  1308.1852}}].

\bibitem{Kaiser:2017hqn}
N.~Kaiser, \emph{{Why there is no Newtonian backreaction}},
  \href{http://dx.doi.org/10.1093/mnras/stx907}{\emph{Mon. Not. Roy. Astron.
  Soc.} {\bf 469} (2017) 744--748},
  [\href{http://arxiv.org/abs/1703.08809}{{\tt 1703.08809}}].

\end{mcitethebibliography}\endgroup

\end{document}